\documentstyle{article}
\begin{document}
\centerline{\bf EXTERIOR DIFFERENTIAL FORMS}
\centerline{\bf IN FIELD THEORY}

\bigskip
\centerline {\it L.~I. Petrova }
\bigskip
\centerline{{\it Moscow State University, Department of Computational }}
\centerline{{\it Mathematics and Cybernetics}}
\centerline{{\it 119899, Moscow, Russia}}
\centerline{E-mail:ptr@cs.msu.su}

\begin{abstract}
A role of the exterior differential forms in field theory is connected with a fact 
that they reflect properties of the conservation laws. In field theory a role of the closed 
exterior forms  is well known. A condition of closure of the form means 
that the closed form is the conservative quantity, and this corresponds to {\bf the 
conservation laws for physical fields}.
In the present work a role in field theory of the exterior forms, which correspond 
to {\bf the conservation laws for the material systems} is clarified. These forms are defined on the 
accompanying nondifferentiable manifolds, and hense, they are not closed. 
Transition from the forms, which correspond to the conservation laws for the 
material systems, to those, which correspond to the conservation laws for 
physical fields (it is possible under the degenerate transform), describe a mechanism 
of origin of the physical structures that format physical fields. In the work it is shown that 
the physical structures are generated by the material systems in the evolutionary 
process. In Appendices we give an analysis of the principles of thermodinamics
and equations of the electromagnetic field. A role of the conservation laws in
formation of the pseudometric and metric spaces is also shown.
\end{abstract}

{\bf Introduction}. A specific feature of the exterior differential forms is that 
at the same time they possess algebraic as well as geometric, and topologic, 
and differential, and integral, and many other properties. It is explained 
by their complicated internal structure (homogeneity with respect to the basis, 
skew symmetry, the integration of elements which are composed of two objects 
with different nature: algebraic (coefficients of the form) and  geometric 
(components of the basis) ones, a structure connection between the forms 
of different degrees, a dependence on space dimension and on topology of 
the manifold. Under a conjugation of the form elements, objects of every elements, 
forms of different degrees, exterior and dual ones, an so on, there realized 
invariant and structure properties  of the exterior differential forms. 
Just these properties are essential for the invariant field theory. They correspond 
to the conservation laws and enalbes one to describe a variety of the physical 
structures which constitute physical fields.

The closed forms have invariant and structure properties. They correspond to the conservation laws 
for physical fields. A role of the closed forms has been described in works by Willer, Shuts and other 
authors [1-3].In the present work a role in field theory of the forms, which are not closed 
because they are defined on the arbitrary nondifferentiable manifolds. Such forms arise under 
a description of the conservation laws for the physical systems. (About a role 
of the closed forms in the existing field theories it  
will be shortly said in the Appendix 1).

Field theories are based on the conservation laws. The exterior differential forms 
enable one to study properties and specific features of the conservation laws 
and thus to disclose the basis of field theory.

At present there are many problems associated with 
the conservation laws. An approach to the conservation laws, their 
mathematical expression and physical treatment turn out to be different 
in different branches of science. A concept "the conservation law" 
in different branches of science carry different sense. 
In physics the conservation laws reveal themselves as conservative quantities 
(such conservation laws may be called  exact ones); in mechanics of continuous 
media these conservation laws establish a balance between a change of 
physical quantities (energy, linear momentum, angular momentum, and mass) 
and relevant external forcing (such conservation laws may be called  
the balance conservation laws); in thermodynamics the conservation laws prove  
to be relevant to the principles of thermodynamics. And what have they in common?

The exact conservation laws relate to {\bf physical fields}. $\{$Physical fields 
(electromagnetic, gravitational, nuclear and etc)
[4] are special forms of the matter which are carriers of interactions$\}$. 
The balance conservation laws are those for {\bf material systems}. $\{$A material 
system is a variety of elements which have internal structure and interact 
to one another. Examples of elements that constitute the material system 
are electrons, protons, neutrons, atoms, fluid particles, cosmic objects and 
others. As examples of material systems it may be thermodynamic, 
gas dynamical, cosmic systems, systems of elementary particles (pointed above) 
and others. The physical vacuum in its properties may be regarded as an analogue 
of the material system that generates some physical fields$\}$. As it will be 
shown in Appendix, the principles of thermodynamics are not special balance 
conservation laws. They combine the balance conservation laws for energy and 
linear momentum, and it enables one to understand a nature of their 
interactions.

In the present work we show that the exact conservation laws follow from the 
balance conservation laws. And {\bf an interaction of the 
balance conservation laws that appear to be noncommutative} plays a crucial 
role. $\{$The principles of thermodynamics may be regarded as an example of 
accounting for this interaction in the thermodynamic systems$\}$. A transition from 
the balance conservation laws to the exact ones is accompanied by origin of 
the physical structures and it forms the basis of the evolutionary processes [5-10].
It is evident that material systems generate the physical structures, and such 
structures format the physical fields. 

In section 1 some properties of the exterior forms utilized 
in the work are presented. In section 2 it is shown that the closed forms 
may correspond to the exact conservation laws. $\{$For the sake of clearness  
in sections 2-8 we introduce a double notation. By simple lettering and by italic 
one we respectively designate a name, which has a physical sense, and that, 
which explains the mathematical sense$\}$.  In the next sections we demonstrate a 
noncommutativity of the balance conservation laws, establish a relation 
between the balance conservation laws and exact those, and disclose a 
mechanism of origin of the physical structures that produce the physical fields. 

In Appendix 1 we show that the invariant and metric properties of the 
closed exterior differential forms constitute the basis of existing field 
theories. In Appendix 2 we give an analysis of the principles of 
thermodynamics that disclose an interaction of the balance conservation 
laws for energy and for linear momentum. In Appendix 3 
it was shown an influence of the noncommutativity of the balance conservation 
laws on a development of instability. In Appendix 4 equations of the electromagnetic 
field were analyzed. In Appendix 5 we present a table of interactions. In 
Appendix 6 a role of the conservation laws in formation of the pseudometric and 
metric spaces is shown, and an analysis of the Riemann space and the Einstein 
equation is presented. In Appendix 7 functional properties of solutions to 
the differential equations, that are essential for the field theory, are 
analyzed. These properties reflect those of the conservation laws.

\bigskip
{\bf 1. Some properties of exterior differential forms.} An exterior differen\-tial 
form of degree p (p-form) may be presented as [1-3,11-14]
$$\theta ^p\,=\,A_{\alpha _1...\alpha _p}\,dx^{\alpha _1}\,dx^{\alpha _2}\,...\,
dx^{\alpha _p}, \quad 0\leq p\leq n \eqno(1)$$
where the basic $dx^{\alpha }$, $dx^{\alpha }\,dx^{\beta }$, $dx^{\alpha }\,
dx^{\beta }\,dx^{\gamma },\,...$ 
obey the condition
$$dx^{\alpha }\,dx^{\alpha }\,=\,0 $$
$$ dx^{\alpha }\,dx^{\beta }\,=\,-dx^{\beta }\,dx^{\alpha }\quad\alpha \neq \beta\eqno(2)$$

A differential (exterior) of the form $\theta ^p$ is expressed by the formula
$$d\theta ^p\,=\,dA_{\alpha_ 1\,...\alpha _p}\,dx^{\alpha _1}\,dx^{\alpha _2}
\,...dx^{\alpha _p}\eqno(3)$$
and it proves to be the differential form of degree $(p+1)$.

We will point out some properties of the closed forms that correspond 
to the exact conservation laws.

1) A closure condition of the $p$-form $\theta ^p$ (form of degree $p$) is written as
$$d\theta ^p\,=\,0\eqno(4)$$
(it is obvious that the closed form is a conservative quantity).

2) If a form is closed only on some structure, i.e. on pseudostructure, then 
the closure condition may be written in the form 
$$d_{\pi }\theta ^p\,=\,0\eqno(5)$$
where the pseudostructure $\pi$ obeys the condition
$$d_{\pi }\,^*\theta ^p\,=\,0\eqno(6)$$
here $^*\theta ^p$ is the dual form. (One can see that a form closed on 
pseudostructure is a preserving object). 
$\{$To the exterior differental form on the differentiable 
manifold there corresponds the skew-symmetric tensor. To 
the dual form that describes the pseudostructure there 
corresponds the pseudotensor dual to the skew-symmetric 
tensor. The pseudostructures format cohomologies 
(cohomologies by De Rham, singular cohomologies [11]), sections of the 
cotangent bundles and so on. They correspond to eikonals which 
are on one hand the level surfaces and on the other the cut 
surfaces$\}$. 

3) A form is called exact one if it is equal to total differential:
$$\theta ^p\,=\,d\theta ^{p-1}\eqno(7)$$
The exact forms are closed identically: $d\theta ^p\,=\,d\,d\theta ^{p-1}\,=\,0$

4) Any closed form is a differential. An exact form is a total differential. 
A closed inexact form is an interior one on the pseudostructure differential.
$$\theta ^p\,=\,d_{\pi }\theta ^{p-1}\eqno(8)$$

5) From Eqs. (7) and (8) it follows that there exist a connection between forms 
of sequent degrees. There is also a similar integral relation
$$\int _{c^{p+1}}\,d\theta ^p\,=\,\int _{\partial c^{p+1}} \theta ^p$$

The theorems by Stokes and Gauss are special cases of this relation.

From the definition of the form it follows that elements of differential 
of the form are equal to components of their 
commutator. Thus, if the form of first degree is expressed as $\theta\,=\,a_{\mu }\,d\xi^{\mu }$, 
then $d\theta=K_{\alpha \beta }\,d\xi ^{\alpha }\,d\xi ^{\beta }$, where 
components of the form commutator are $K_{\alpha \beta }=a_{\beta ;\alpha }-
a_{\alpha ;\beta }$. Here $a_{\beta ;\alpha }$, $a_{\alpha ;\beta }$ are the covariant 
derivatives. In the case of differentiable manifold the covariant derivatives 
coinside with ordinary ones and the commutator components can be written as
$$K_{\alpha \beta }\,=\, \left({{\partial a_{\beta }}\over {\partial \xi ^{\alpha }}}
\,-\,{{\partial a_{\alpha }}\over {\partial \xi ^{\beta }}}\right)
\eqno(9)$$

If a form is defined on the nondifferentiablle manifold, then  
an additional term will appear in the commutator, this term is a commutator 
of the metric form of manifold.

\bigskip
{\bf 2. Conservation laws for physical fields.}({\it A closure condition 
of exterior forms}). As it is seen from the closure conditions (4), (5), 
a closed form is a conservative quantity, and hence it may correspond to 
the exact conservation law. And the closed inexact form is a conservative 
object, namely, it is a conservative quantity only on some pseudostructure $\pi $. 
The closure conditions for the exterior form ($d_{\pi }\,\theta ^p\,=\,0$) 
and the dual form ($d_{\pi }\,^*\theta ^p\,=\,0$) are 
mathemati\-cal expression of the exact conservation law.

The exact conservation laws correspond to physical structures that format 
physical fields. Preserving objects, e.g. conservative quantities (closed 
exterior forms), on the pseudostructures (dual forms) are the physical 
structures that format physical fields:
$$
\def\\{\vphantom{d_\pi}}
\cases{d_\pi \theta^p=0\cr d_\pi {}^{*\mskip-2mu}\theta^p=0\cr}\quad
\mapsto\quad
\cases{\\\theta^p\cr \\{}^{*\mskip-2mu}\theta^p\cr}\quad\hbox{---}\quad
\hbox{physical structures}\quad\mapsto\quad\hbox{physical fields}
$$              
Equations for the physical structures ($d_{\pi }\,\theta ^p\,=\,0$, 
$d_{\pi }\,^*\theta ^p\,=\,0$) turn out to coincide with a mathematical 
expression of the exact conservation law. It is seen that the exact 
conservation law is that for physical fields.

As any closed form is a differential (either total if the form is exact 
one: $\theta ^p\,=\,d\theta ^{p-1}$, or interior on the pseudostructure: 
$\theta ^p\,=\,d_{\pi }\,\theta ^{p-1}$ if the form is inexact) of the form 
of lower degree, then the form of lower degree may correspond to a potential.
And the form degree indicates a type of potential. The potential 
is a scalar if $l=0$ ($l=p-1$), it is a vector if $l=1$, and it is a tensor 
if $l=2,3$. The closed inexact forms of zero, first and second degrees relate 
respectively to the pseudoscalar, pseudovector and pseudotensor (vortex) fields.
$\{$As it was pointed before, the closed form is a conservative quantity. On the 
other hand, as it is  a differential of some form, which in this case plays 
a role of the potential, the closed form reveals as a potential force. That 
is, the closed form is dual object. This duality of the 
closed forms discloses properties of the physical fields decribed as the 
carriers of interactions. Below it will be shown, with respect to what the 
closed form reveals as a potential force.$\}$
 
Thus, an application of the closed exterior differential forms enable one 
to see a relation between the conservation laws and the physical structures 
that produce physical fields. It may be shown that invariant and metric 
properties  of the closed exterior differential forms that correspond to 
the exact conservation laws constitute the basis of the existing theories 
which describe physical fields. We are able to verify that all existing 
theories are complemented by additional conditions of invariance or covariance 
which are the closure conditions for exterior or dual forms. (In more details 
this subject is described in Appendix 1). 

The existing field theories that are based on the exact conservation laws 
allow to describe physical fields, find possible physical structures, and 
understand a variety of physical fields. However, such invariant theories 
cannot give an answer to  the question concerning a mechanism of the genesis 
of physical structures. Only a theory that does not base on the closure 
conditions for forms can give an answer. These conditions have to be achieved 
by themselves spontaneously. A realization of these conditions 
implies a formation of the closed form, and this corresponds to the 
conservation law and gives an indication of a production of physical structure. 
It is evident that only the evolutionary theory can answer the question about a mechanism of generation of the physical 
structures. Such a theory is one based on the balance conservation laws for the material 
systems. (It also bases on the properties of exterior forms, however, in this case 
the exterior form are defined on the nondifferentiable manifold, and therefore they are 
nonclosed. These forms are nonintegrable ones. Topological properties of commutators 
for such forms enable one to understand a role of the balance conservation laws in 
evolution processes and in formation of physical fields.

\bigskip

{\bf 3. Noncommutativity of the balance conservation laws for material systems}. 
{\it (Nonidentity of the evolutionary relation obtained from the balance conservation 
laws)}. The conservation laws for the material system (continuous medium) 
express the following: 
a change of any physical quantity in some volume for a given time interval is 
balanced out by the flux of this quantity across the boundary of the volume 
and by the source actions. Under transition to the differntial expression the fluxes 
are substituted by divergences. And the differential or integral equation 
is supplemented with a dependence of the physical quantities on the state function 
of the material system. Below a relation (in the differential 
forms) for the state function will be obtained. Just this relation disclose a 
specifics of the evolutionary processes. It should be underlined once more that 
the conservation laws for the material medium are balance ones.

The balance conservation laws (for energy, linear momentum, angular momentum, 
and mass) establish a balance between a change of physical quantities and 
an action on the system. And every balance conservation law depends on 
relevant actions (so the conservation law for energy depends on energetic 
actions, the conservation law for linear momentum does on force ones, etc). 
In real processes such actions have a different nature. Therefore the 
balance conservation laws prove to be noncommutative. $\{$What is the 
``noncommutativity" of the balance conservation laws?  Suppose, that first 
the energetic and then the force perturbations act on a local domain of the 
material system (element and its neighborhood). And let at the initial moment 
the local domain be in some state ${\bf A}$. According to the balance 
conservation law for energy, under an influence of the energetic perturbation 
the local domain develops from the state ${\bf A}$ into any state ${\bf B}$. 
Then according to the balance conservation law for momentum under an influence 
of the force perturbation it develops from the state ${\bf B}$ into any state 
${\bf C}$. Suppose now that the sequence of the actions changes, namely, first 
the force perturbation and then the energetic one act, and the system develops 
first into any state ${\bf B}^*$ and then does into the state ${\bf C}^*$. 
If the state ${\bf C}^*$ coincides with the state ${\bf C}$, that is, the result 
does not depend on a sequence of perturbation of different type (and on a sequence 
of performing the relevant balance conservation laws), then it means that the 
balance conservation laws commutate. If the state ${\bf C}^*$ does not coincide 
with the state ${\bf C}$, then it means that the balance conservation laws 
prove to be noncommutative$\}$. This is just of decisive importance for 
evolutionary processes and a mechanism of origin of the physical structures. 
{Because of noncommutativity of the balance conservation laws, an external forcing 
that experience the material system cannot directly convert into physical 
quantities (energy, linear momentum, angular momentum, mass) of the system, but 
they convert into some quantity that acts as internal force and is a cause of 
the evolutionary processes}.

To understand how the balance conservation laws interact to one another it 
is necessary to study a conjunction (self-consistence) of equations governing these 
laws. 

The balance conservation laws may be described by means of differential 
equations [15-18]. If the material system is not dynamical one (as in the case 
of thermodynamic system), then the equations of the balance conservation 
laws may be written in terms of increments of physical quantities and 
governing variables.

Equations are conjugate ones if they may be contracted into identical 
relations for differential, i.e. for a closed form. Let analyze the equations 
that describe the balance conservation laws for energy and linear momentum. 

We introduce two frames of reference: the first is inertial one 
(this system is not connected with the material system), and the second 
is accompanying one (this system is connected with a manifold created by 
trajectories of elements of the material system). The energy equation 
in the inertial frame of reference may be reduced to the form:
$${{D\,\psi}\over {D\,t}}\,=\,A \eqno(10)$$
where $D/D\,t$ is the total derivative with respect to time, $\psi $ is the functional 
of the state that specifies the material system, $A$ is a quantity, which  depends on 
specific features of the system and on external energy actions on the 
system. {The action functional, entropy, 
wave function may be regarded as examples of the functional $\psi $. Thus, the equation 
for energy represented in terms of the action functional $S$ has a similar form: 
$DS/Dt\,=\,L$, where $\psi \,=\,S$, $A\,=\,L$ is the Lagrange function. In mechanics of continuous media the equation for 
energy of ideal gases may be presented in the form [16]: $Ds/Dt\,=\,0$, where 
$s$ is entropy. In this case $\psi \,=\,s$, $A\,=\,0$. It is worth note that the examples presented show 
the mutual relation between the action functional and entropy}.

In the accompanying frame of reference a total derivative with respect to time 
transforms into that along the trajectory. The equation (10) turns out to be written 
in the form
$$ {{\partial \psi }\over {\partial \xi ^1}}\,=\,A_1 \eqno(11)$$
here $\xi ^1$ is a coordinate along the trajectory. In a similar manner, in the 
accompanying frame of reference the equation for linear momentum appears 
to be reduced to the equation of the form
$${{\partial \psi }\over {\partial \xi ^{\nu }}}\,=\,A_{\nu },\quad \nu \,=\,2,\,...\eqno(12)$$
where $\xi ^{\nu }$ are coordinates in the direction being normal to the trajectory, 
$A_{\nu }$ are the quantities that depend on specific features of the system and external 
force actions.

The Eqs. (11), (12) may be convoluted into the relation
$$d\psi\,=\,A_{\mu }\,d\xi ^{\mu },\quad (\mu\,=\,1,\,\nu )\eqno(13)$$
where $d\psi $ is the differential expression $d\psi\,=\,(\partial \psi /\partial \xi ^{\mu })d\xi ^{\mu }$.

The relation (13) may be written in the form:
$$d\psi \,=\,\omega \eqno(14)$$
here $\omega \,=\,A_{\mu }\,d\xi ^{\mu }$ is the differential form of the first degree.

As the balance conservation laws are evolutionary ones then the relation obtained 
is also evolutionary law. {The proper evolution relation corresponds to 
every material system (see Appendices (2)-(4)$\}$.

The relation (14) has been obtained from the balance conservation laws for 
energy and linear momentum. In this context the form $\omega $ is that of the 
first degree. If the equations of the balance conservation laws for 
angular momentum be added to the equations for energy and linear momentum, 
then in the evolutionary relation this form will be that of the second degree. 
And in a combination with the equation of the balance conservation law 
for mass this form will be a form of degree 3. Thus, in the general case 
the evolutionary relation may be written in the form

$$d\psi \,=\,\omega ^p$$
where the degree of the form $p$ takes the values $p\,=\,0,1,2,3$.. (The evolutionary 
relation for $p\,=\,0$ is similar to that in the differential forms, and it has been 
obtained from interaction of energy and time or momentum and coordinate.) 

In the left-hand side of the evolutionary equation there is the functional 
expression $d\psi $, which specifies a state of the material system, and 
in the right-hand side there is the form $\omega ^p$ which coefficients 
depend on external actions. The meaning of the evolutionary relation 
lies in a fact that it discloses a specific feature of dependence of the 
material system state on external actions. As it will be shown below, 
the evorutionary relation has a specific feature (it may be nonidentic one) which 
enables one to determine a mechanism of transition of the material system 
from nonequilibrum state to the equilibrum or locally equilibrum one, and this 
has a decisive importance for the evolutionary process. $\{$It should be 
pointed out the following. If to a state of the material system there correspond 
a differential of any function, then this indicates that a state of the material 
system is the equilibrum or locally equilibrum one. And if the differential 
is absent, then this means that the system is in the nonequilibrum state. 
As it will be shown below, owing to this specific feature the evolutionary 
relation enables one 
to detect a presence or absence of the differential and as the result to classify 
a state of the system$\}$. 

At this point we show that for real processes {\bf the evolutionary relation appears to be 
nonidentical one }. A relation may be 
identical one if it relates any measurable metric or invariant objects, 
i.e. the objects which may be compared. $\{$A concept ``nonidentic relation" may be seemed as contradictive one. However, 
this concept carries an in-depth meaning. The identic relation establishes an 
exact correspondence between quantities entering into that. The nonidentic relation 
can establish an exact correspondence between quantities entering into that 
only under some supplementary conditions. If these supplementary conditions do 
not satisfied, this relation has a physical meaning as well. If this relation is 
evolutionary one, then it proves to be a selfvariating relation, that is, a variation 
of one object of the relation forces a change of the other object, and in 
turn a change of the second object leads to a change of the first one 
and so on. As one object is nonmeasurable object, than the other object cannot be 
compared with the first one, and hense a process of selfvariation cannot be 
terminated without additional conditions. Additional conditions can be realized by 
themselves under a selfvariation of the nonidentic relation owing to any
degrees of freedom. Just the nonidentic relations, which the evolutionary relations belong to, 
can describe a self-organization of the material systems. The principle of 
self-organization will be clarified later$\}$.

Let as examine the relation (14). For real processes the form $\omega $ that 
stands in the evolutionary relation (14) and depends on the external actions 
appears to be nonclosed and hence cannot be an invariant object. For the 
form to be closed it needs the differential of the form or its commutator 
be zero (elements of a differential of the form are equal to components of 
its commutator). Let us consider a commutator of the form $\omega \,=\,A_{\mu }d\xi ^{\mu }$. Components 
of the commutator may be written as follows:
$$K_{\alpha \beta }\,=\,\left ({{\partial A_{\beta }}\over {\partial \xi ^{\alpha }}}\,-\,
{{\partial A_{\alpha }}\over {\partial \xi ^{\beta }}}\right )\eqno(15)$$
(here a term that is connected with a nondifferentiability of the manifold was 
not taken into account as yet). Coefficients of the form $\omega $ have been obtained either 
from the balance conservation law for energy or from that for linear momentum. 
It means that in the first case the coefficients depend on the energetic 
action and in the second case they depend on the force action. In actual 
processes energetic and force actions have a different nature and appear 
to be nonconsistent. The commutator of the form $\omega $ constructed from derivatives 
of such coefficients is nonzero. This means that a differential of the form $\omega $ 
is nonzero as well. Thus, the form {} proves to be nonclosed and isn't an 
invariant object. Therefore, in the evolutionary relation there in 
noninvariant term. Such a relation cannot be identical one. Hence,  
without a knowledge of a particular expression for the form {}, one can argue 
that for an actual processes the evolutionary relation proves to be 
nonidentical because of nonconsistency of the external action. To emphasize 
this fact, it is reasonable to write down the relation (14) in the form 
$$d\psi\, \cong \,\omega \eqno(16)$$
(it was introduced the sign $\cong $ instead of the equality sign $=$). In a similar 
manner one may prove the nonidentity  of the general evolutionary relation 
and to write it in the form:
$$d\psi \,\cong \,\omega ^p \eqno(17)$$
A nonidentity of the evolutionary relation means that equations of the 
balance conservation laws turn out to be nonconjugate (thus, if from
the energy equation to obtain a derivative of $\psi $ in the direction 
along the trajectory and from the momentum equation to find a derivative 
of $\psi $ in the direction normal to the trajectory and next to calculate their 
mixed derivatives, then from the condition that the commutator of the form $\omega $ 
is nonzero it follows that mixed derivatives prove to be noncommutative). 
It points that these balance conservation laws don't commutate.

\bigskip
{\bf 4. Nonequilibrum property of material system. Selfvariation of 
the state of material system.} {\it Topologic properties of commutator 
for nonintegrable form}. What is follow from noncommutativity of the 
balance conservation laws? Let us analyze the evolutionary relation (14). 
If the balance conservation laws for energy and momentum be commutative, 
then the relation (14) turns out to be identic, and the differential $d\psi $ can 
be obtained from that. It may be regarded that $\psi $ is a function 
of the state. And an existence of the function of state means that the state 
is locally balanced one. This means that external actions directly transform 
into physical measurable quantities of the system, namely, energy and momentum 
of elements of the material system. But because of the evolutionary relation 
is not identical, one cannot obtain the differential $d\psi $ from that relation, 
and this means that there is no the function of state. The state of system is nonequilibrum 
one. Because of the noncommutativity of the balance conservation laws, the 
external forces cannot directly transform into measurable quantities of the system, 
and they transform into any nonmeasurable quantity. This nonmeasurable 
quantity, which is described by the commutator of the form $\omega $, acts as
internal force. This means that because of the noncommutativity of the balance 
conservation laws  the state of material system proves to be nonequilibrum one 
and there is an internal force described by the commutator of the 
nonintegrable form $\omega $. In the general case this nonmeasurable quantity is 
described by the commutator of the nonintegrable form $\omega ^p$. The 
nonequilibrum is a moving force of the evolutionary process. A further 
analysis of the evolutionary relation and the commutator of the form $\omega ^p$ 
allows to understand a mechanism of the evolutionary process. 

A nonidentical relation, if it is a evolutionary relation, appears to be 
selfvariating: a change of any object of the relation forces a change of other 
object and in turn a change of the second object leads to a change of the 
first one and so on. As one of objects of the nonidentical relation is 
nonmeasurable object, then the other object cannot be exactly compared with
the first object. This property of the nonidentical evolutionary relation 
makes it possible to describe a selfvariation  of the nonequilibrum state 
of  the material system. In this case topologic properties of the commutator 
of nonintegrable form (this form is defined on the nonintegrable manifold) 
play a deciding role.

If manifold, on which the exterior form is defined, is nondifferentiable one, 
then an additional term that contains a commutator of the metric form of 
manifold (this commutator specifies a deformation of the manifold) will 
enter into a commutator of the exterior form. The form $\omega ^p$, that stands in 
the evolutionary relation, is defined on the accompanying manifold, which for 
actual processes turns out to be nondifferentiable one as it is formatted 
contemporaneously with a change of the state of material system and depends 
on the physical processes. Hence, a term containing the characteristics of the 
manifold will enter into a commutator of the form $\omega ^p$ in addition to a term, 
which is connected with derivatives of coefficients of the form. An interaction 
between these terms of different nature just describes a mutual change of 
the state of material system.

Let us examine this with an example of commutator of the form 
$\omega \,=\,A_{\mu }\,d\xi ^{\mu }$ that enters into 
the evolutionary relation (14). We assume that at the beginning an associated 
manifold was differentiable. In this case a commutator of the form $\omega $ can 
be written in the form (15). If at the next point in time any action effects 
on the material system, then this commutator turns out to be nonzero. A state 
of the material system becomes nonequilibrum and it will appear an internal 
force whose action will lead to a deformation of the accompanying manifold.
Then the accompanying manifold fails to be differentiable. In a commutator
of the form $\omega $ it will appear an additional term, that specifies a deformation 
of the manifold and is a commutator of the metric form of manifold. {If it is 
possible to define the coefficients of connectivity $\Gamma _{\alpha \beta }^{\sigma }$ 
(for nondifferentiable manifold 
they are asymmetric ones), then a commutator of the form may be written as
$$K_{\alpha \beta }\,=\,\left ({{\partial A_{\beta }}\over {\partial \xi ^{\alpha }}}\,-\,
{{\partial A_{\alpha }}\over {\partial \xi ^{\beta }}}\right )\,+\,(\Gamma _{\beta \alpha }^{\sigma }\,-\,
\Gamma _{\alpha \beta }^{\sigma })\,A_{\sigma }\eqno(18)$$
where $(\Gamma _{\beta \alpha }^{\sigma }\,-\,
\Gamma _{\alpha \beta }^{\sigma })$ is a commutator of the metric form (which specifies a torsion of 
manifold)}. An emergence of the second term can only change a commutator and cannot 
make it zero (because terms of the commutator have different nature). In the 
material system the internal force will continue to act even without external 
actions. The further deformation (torsion) will go on. This leads to a change of 
commutator of the metric form, produces a change of the exterior form and its 
commutator and so on. A process of selfvariation of the commutator that governs 
by nonidentic evolutionary relation specifies a change of the external force 
and selfvariation of nonequilibrum state of the material system. At this point it 
should be emphasized that such selfvariation of the state of material system 
proceeds under an action of internal (rather then external) forces and may go on 
even without action of external forces. And a state of material system remains 
nonequilibrum, and in actual physical process an internal force  could give 
rise to development of instability in the material system. $\{$For example, this 
was pointed out in works by Prigogine [19]. "The excess entropy" in his 
works is analogous to a commutator of nonintegrable form for thermodynamic system. 
"A production of excess entropy" leads to 
development of instability$\}$. An internal force cannot continuously become 
equal to zero. The material system cannot continuously transform into an 
equilibrum (without internal forces) state. However, in the material system 
an equilibrum state may be locally realized if internal forces transform 
into potential ones. As an analyses of the evolutionary equation shows, 
it is possible under additional conditions and it corresponds to emergence 
of the physical structure.

\bigskip
{\bf 5. A mechanism of origin of the physical structure.} ({\it Degenerate 
transformation}). For actual processes a differential of the form $\omega ^p$ that 
enters into the evolutionary equation (as well as the commutator) is 
nonzero:
$$
d\,\omega ^p\,\neq \,0\eqno(19)
$$
i.e. the form $\omega ^p$ is nonclosed. To the physical structure one has to assign 
a closed form on pseudostructure, that is to say, that to the physical 
structure there corresponds some differential $d_{\pi }\psi$. Such a differential 
may be obtained from the evolutionary relation only if the form $\omega ^p$ 
is closed inexact one, in other words, if the form is subjected to the conditions  
$$
\cases{d_{\pi }\omega ^p=0\cr
d_{\pi }\,^*\omega ^p=0\cr} \eqno(20)
$$
(in this case the form appears to be an interior differential of any form, namely 
$\omega ^p\,=\,d_{\pi} \vartheta $. The evolutionary relation becomes identical on the 
pseudostructure $\pi$, and a differential $d_{\pi}\,\psi\,=\,d_{\pi }\,\vartheta $, that  
corresponds to the physical structure, can be determined from that relation). 

A transition from the condition (19) to the condition (20) that corresponds
to origin of the physical structure (and a transformation of internal 
forces into potential ones) is possible only as {\bf the degenerate transform}, 
i.e. it is a transition which does not preserve a differential. $\{$At this point 
it should be underlined that in this case the degenerate transform is realized
as a transition from the accompanying frame of reference to the inertial 
one$\}$. Some additional condition has to correspond to the degenerate transform. 
(A vanishing of such functional expressions as Jacobians, determinants, the 
Poisson brackets, and others may be regarded as examples of such conditions). 
Because the evolutionary relation describes the material systems and 
coefficients of the form depend on properties of the material system, 
then it is evident that a condition of the degenerate transform has to be 
caused by properties of the material system. As an example, the system 
may have any degrees of freedom. Namely, under realization of any degree 
of freedom it may take place a redistribution between physical quantities 
(for example, between energy and linear momentum) in such a way that they 
become measurable simultaneously . Such degrees of freedom may be 
translational ones, internal degrees of freedom of elements of the system 
and so on.

Thus, the physical structure can arise, if, firstly, the material system 
undergoes nonconsistent external actions and {\it a commutator of the form $\omega ^p$ 
is nonzero}. And, secondly, if the material system has degrees of freedom, 
i.e. {\it there are conjugation conditions for the exterior form} $\omega ^p$. 
However, even if these conditions are satisfied, the physical structure appears 
only in the case, when degrees of freedom of the system are realized (in 
physical process), namely, {\it the conditions of conjugation are satisfied}. It may 
take place under selfvariation of the state of material system. And it may 
be realized only spontaneously because it is caused by internal (rather then 
external) reasons (degrees of freedom are characteristics of the material 
system and not of external actions).

A creation of the physical structure is a transition of a quantity that 
acts as internal force into a measurable quantity that acts in the direction normal 
to the pseudostructure as potential 
force . $\{$Above it was poined out the duality 
of the closed form (as conservative quantity and as potential force). This duality has a 
physical meaning. The closed form as conservative quantity relates to the physical field. 
The object, which is conservative quantity on the pseudostructure, is an element 
of physical field. And the closed form reveals as potential force with respect to 
the material system. The potential force is an action of the formation 
originated (see below) on elements of the material system. The 
potential forces are described, for example, by jumps of derivatives in the 
direction normal to the characteristics, to the potential surfaces, and so on.$\}$  

The physical structure and some measurable quantity that acts as 
potential force reveal as a new measurable and observable formation that 
spontaneously arises in the material system. $\{$Fluctuations, pulsations, 
waves, vortices, massless particles are examples of such formations$\}$. 
In the physical process this formation spontaneously extracts from the local 
domain of material system and so it allows the local domain of material 
system to get rid of internal force and come into locally balanced state. 
A formation, that has been created in some local domain of the material system 
and liberated from that, begins to act on the neighboring local domain 
as a potential force (this forcing was created by system in itself, 
and therefore this is potential forcing (rather then arbitrary one). 
The neighboring domain of material system works over this action, 
which appears to be external for it. If in the process the conditions 
of conjugation turn out to be satisfied again, then the neighboring domain 
create a formation of its own. In such a way the formation can move 
relative to material system. A velocity of moving relative to material 
system is not a parameter of the system, but this is some quantity that 
is realized at every time as the conjugation condition. If the material 
system is homogeneous one, then a velocity will have the same values 
(but this is not a constant because it anew arise at every point of time 
of the evolutionary process). The speed of sound, the speed of 
electromagnetic waves, the speed of light are formatted in such a way. It is 
evident, that a velocity of moving relative to material system is defined 
by internal properties of the system. (As example, the speed of sound 
$a$:$\, a^2=(d\,p/d\,\rho)_s$ , where the pressure $p$ and density $\rho $ 
are characteristics of the material system and the subscript $s$ means 
a constancy of entropy).

\bigskip
{\bf 6. Characteristics of a formation originated: intensity, spin, absolute and 
relative speeds of propagation of the formation}. {\it (Characteristics of exterior 
and dual forms, value of commutator of the nonintegrable form, properties 
of the material system)}. As a formation originated is a result of transition 
of nonmeasurable quantity, which is described by the commutator (and acts like 
internal force), into measurable quantity (potential force), then it is evident 
that an intensity of the formation originated is defined by a quantity, which 
was stored by the commutator of the nonintegrable form at the moment of its 
appearing. And the first term of the commutator that formatted by mixed  
derivatives of the form coefficients governs an intensity of the formation, 
whereas the second term that specifies a deformation of the accompanying manifold 
(bending, torsion, curvature) is fixed as any internal characteristics of the formation 
originated. Spin is an example of such a characteristics, and a value of spin 
depends on degree of the form. An integrating direction, i.e. pseudostructure 
that is defined by the dual form, determines an absolute speed of propagation of the 
formation originated (it is a speed in the inertial frame of reference). The speed 
of propagation relative to the material system (a speed in the accompanying 
frame of reference), as examples of that are speed of sound, speed of light, speed of electromagnetic 
waves and so on, is defined by the conjugation conditions, i.e. by degrees of freedom 
of the material system.

In such a way the following correspondence is established:

1) an intensity of the formation (potential force) is {\it a value of the first 
term in the commutator of nonintegrable form} at the instance of formation 
creation;

2) spin is {\it the second term in the commutator that is connected with the 
metric commutator};

3) a preserving quantity (a charge) is {\it the exterior form that has been realized}; 

4) an absolute speed of propagation of the formation arisen (a speed in the inertial 
frame of reference) is the integrating direction--pseudostructure--{\it dual form};

5) a speed of propagation relative to the material system (the speed of sound, 
the speed of light, the speed of electromagnetic wave) is a speed in the accompanying 
frame of reference--conjugation conditions--{\it  additional conditions connected with 
properties of the material system}.

\bigskip
{\bf 7. A formation of pseudometric and metric spaces.} ({\it An integration 
of the nonidentical evolutionary relation}). An analyses of integrability of 
the nonidentical evolutionary relation explains a process of formation of 
pseudostructures and thereby make more evident a mechanism of formation of 
pseudometric and metric space.

As it is known, the closed form is a differential (exterior or interior) of 
the form of one less degree. This connection enables one to carry out an 
integration of the closed form and proceed to the form of one less degree.  
Such transitions are possible only in identical relations. It may be shown 
that an integration and transitions with lowering the form degree are allowed 
in nonidentical relations (nonintegrable forms) as well but only in the case 
of degenerate transforms. Under a degenerate transform on the pseudostructure 
it may be obtained the identical relation that can be integrated and it enables 
one to get a relation with the forms of one less degree. The relation obtained 
turns out to be nonidentical as well. By integration (under realization of 
relevant degenerate transform) the nonidentical evolutionary relation with 
forms of degree $p$, one may successively obtain nonidentical relations with 
forms of degree $k$, where $k$ takes values from $p$ to 0. At each transition 
the closed forms on the pseudostructure of sequent degrees $k=p,\,k=p-1, ...,
\,k=0$ are formatted, and this indicates a creation of the physical structures 
of relevant type. A transition to the exact form of zero degree corresponds to 
a creation of some element of the material system (massive particle with internal 
structure). $\{$So called ``spontaneous violation of the symmetry" is an example 
of such a transition$\}$.

To a creation of the physical structures we put into correspondence a formation 
of the pseudostructures with a dimension which depends on the space dimension. It 
may be shown that under a generation of closed forms of sequent degrees $k=p,\,
k=p-1,...,\,k=0$ the pseudostructures of the dimensions $(n+1-k)$: $1,...,\,n+1$ 
are obtained (here $n$ is a dimension of the initial inertial space).
$\{$When deriving the evolutionary relation two frames of reference were 
used. The first system is inertial one which is connected with the space where  
the material system situates and is not directly connected with the material 
system. This is an inertial space, it is the metric space. The second frame of 
reference is proper one, it is connected with accompanying manifold which is not 
metric manifold$\}$. When proceeding to the exact closed form of zero degree 
the metric structure of the dimension $n+1$ is obtained. As a result we get 
that under influence of nonconjugate external forcing (and if there are degrees of freed\-om) 
the material system can transform the initial inertial space of the dimension 
$n$ into a space of the dimension $n+1$. $\{$In the initial space of degree $0$ 
it may be formatted a space of the dimension $1$ (such a space may be a time). 
And in space of the dimension $1$ it may be formatted a space of the dimension $2$ (time and 
space coordinate) and so on. Every material system has their proper time. In 
particular, this approach explains how the proper time is formatted). In the 
initial space of the dimension $3$ it may be formatted a space of the dimension $4$ 
(time and three space coordinates). Such a space can be convoluted and a new 
dimension may happen to be nonrealizable. Thus, the cycle ends and 
a new cycle may begin (this corresponds to that one system may be embedded into the 
other one). A mechanism of formation of the pseudostructures and the metric 
structures can explain, in particular, how the internal construction of elements 
of the material system is formatted$\}$. So it can be seen that the inertial 
spaces are not absolute spaces where actions are developed, but they are 
spaces generated by the material systems. A mechanism of formation of the 
pseudostructures is at the basis of creation 
of the pseudometric spaces and of their transition into the metric spaces. 
(In Appendix 6 a formation of pseudoriemann and 
Riemann's spaces are considered, and conditions of a derivation of the 
Einstein equation are analyzed).

It may be shown that equations of the characteristic surfaces, potential 
surfaces  (simple or double layer), residue equation and others obtained from the 
equations of the mathematical physics are the equations of the 
pseudostructures. As the equations for pseudostructure there serve the 
eikonal equation (the pseudostructure is the level surface, on that the conservative 
physical quantity is definded as it follows from properties of the closed inexact 
form). $\{$In the works [15, 20] it was shown a connection of equations 
for single, double, ... eikonals with the equations of the characteristics, 
Hamilton equations and so on.$\}$ 

\bigskip
{\bf 8. Classification of the physical structures}. ({\it Parameters of 
closed exterior and dual forms}). To obtain the physical structures of 
given physical field it is necessary to consider the material system 
which corresponds to this field. In particular, to obtain the thermodynamic 
structures (fluctuations, phase transitions, etc) one has to analyze the 
evolutionary relation for the thermodynamic systems, to obtain the gas dynamic 
ones (waves, jumps, vortices, pulsations)  he has to employ the evolutionary 
relation for gas dynamic systems, for electromagnetic field he must employ 
a relation obtained from equations for charged particles. Maybe, the physical 
vacuum is an analogue to such material system in the case of elementary particles. 
$\{$Some concrete relations are presented in Appendices (2)-(4)$\}$.

The closed forms that correspond to physical structures are generated by 
the evolutionary relation having parameter $p$ which defines a number of 
interacting balance conservation laws. Therefore, the physical structures 
can be classified by the parameter $p$. The other parameter is a degree 
of the closed forms. As it was shown above, the evolutionary relation of 
power $p$ may generate the closed forms of degree $0\leq k \leq p$. Therefore, 
the physical structures can be classified by the parameter $k$ as well. 
The closed exterior forms of the same degree realized in spaces of 
different dimensions prove to be distinguishable because a dimension of the 
pseudostructures, on which the closed forms are defined, depend on the space 
dimension. As a result, the space dimension also specifies the physical 
structures. This parameter determines properties of the physical structures 
rather then their type.

Hence, from the analyses of the evolutionary relation one can see that 
a type and properties of the physical structures (and accordingly of physical 
fields) for a given material system depend on a number of interacting balance 
conservation laws $p$, on the degree of realized closed forms $k$, and on a 
space dimension. By introduction a classification with respect to $p$, $k$,
and a space dimension we can understand an internal connection of various 
physical fields and interactions (see Appendix 5).

\bigskip
{\bf Conclusions}. Thus, the mathematic tool that bases on properties of 
nonintegrable exterior differential forms enables one to understand a 
role of the conservation laws in the evolutionary processes and disclose 
a mechanism of formation of the physical fields. Because of nonconsistence 
of the external effects the balance conservation laws  
(energy, linear momen\-tum, angular momentum, and mass) for the material systems
prove to be noncommutative ({\it it follows from the nonidentity of the 
evolutionary relation obtained from equations of the balance conservation laws}).
And then, because of noncommutativity of the balance conservation laws, a state 
of the material system turns out to be nonequilibrum, and this is the cause of 
the evolutionary process ({\it a value of the internal force is determined by 
a commutator of the nonintegrable form that enters into the evolutionary 
equation}).  Under some additional conditions that are determined by properties 
of the material system and realized in the physical processes, it is possible 
a conjugation of the balance conservation laws ({\it a degenerate transform 
corresponds to this case}), and it is an indicator of a creation of the physical 
structures (transformation of internal forces into potential ones). And in the 
material system this reveals as arising some measurable formations: fluctuations, 
pulsations, waves, vortices, particles and so on ({\it characteristics of formations 
arisen are determined by ones of the nonintegrable exterior forms, their commutators 
and conjugation condition as well as by characteristics of the closed exterior 
forms and dual forms}).

It was shown that the physical structures which constitutes the physical 
fields are generated by the material systems. This process is accompanied 
with a formation of the pseudometric and metric spaces, and the conservation 
laws govern this processes.

At this point it worth underline that, in spite of these results are 
qualitative ones, they can help while studying the evolutionary processes 
in the concrete material systems and while investigation of concrete 
physical fields and their formations (see Appendices (2)-(6)).

The mathematical theory that explains the evolutionary process in the 
material systems and a mechanism of formation of the physical fields may be 
regarded as the basis of 
the qualitative evolutionary theory of fields. An investigation carried out 
allows to conclude that axioms provided the basis of the existing theories are 
the conjugation conditions of the balance conservation laws for the 
material systems, which generate the physical fields. $\{$It worth emphasize 
that we used a mathematic tool which bases on properties of the {\bf nonintegrable 
exterior differential forms}, i.e. forms defined on the nondifferentiable manifolds. 
A specific property of these forms is a presence of nonidentical relations and 
degenerate transforms (a transition from the accompanying frame of reference 
to the inertial one)$\}$.

\bigskip
\rightline{\bf Appendix 1.}

{\bf Exact conservation laws as the basis of existing field theories}. 
({\it Closed exterior forms in field theory}).

As the closed exterior differential form corresponds to the conservation laws for physical 
fields and potentials, they may be exploited for description of physical fields. 
A dependence of properties of exterior and dual to them forms on a degree of 
the form, space dimension, complex exterior and interior structure of the forms,
various conditions and manners  of conjugation provide a diversity of physical 
structures that format physical fields. It may be shown that essentially 
all physical theories that describe physical fields are based on operators which 
are reduced to operators acting on the closed exterior forms. If in addition to 
the exterior differential $d$ to introduce operators: 1)$\delta$ for an operator  
that takes the form of degree $(p+1)$ to that of degree $p$, 2) $\delta '$ for 
an operator of cotangent transforms, 3)$\Delta $ for the transform $d\,\delta - \delta \,d$,
4) $\Delta '$ for the transform $d\,\delta'-\delta '\,d$, then the operators of 
field theory that act on the differential forms can be written in terms of these 
operators. The operator $\delta $ corresponds to Green's operator, the operator 
$\delta '$ corresponds to the operator of canonical transform, $\Delta$ does to the d'Alembert 
operator in the 4-dimension space, and $\Delta $ corresponds to the Laplace operator 
[13-14]. Eigenvalues of these operators reveal themselves as conjugation conditions for 
elements of the differential forms. In the tensor equations the skew symmetric 
tensors correspond to the closed forms.

All existing field theories are based on the exact conservation laws. The 
essence of these theories are invariant and metric properties  of the closed exterior 
differential forms that correspond to the exact conservation laws. (In 
particular, it may be verified that equations of existing field theories are 
those  that have solutions being the closed or dual to them forms). Into all 
existing field theories there were introduced supplementary conditions of 
invariance or covariance which appears to be the closure conditions for the 
exterior or dual forms. Essentially all existing field theories contain  elements 
of noninvariance, i.e. they are based either on functionals that are not identical 
invariants (such as Lagrangian, action functional, entropy) or on equations 
(differential, integral, tensor, spinor, matrix and so on) that have no 
identical invariance (integrability or covariance). Such elements of 
noninvariance are, for example, nonzero value of the curvature tensor in the 
Einstein theory [21], the indeterminacy principle in the Heisenberg theory, 
the torsion in the theory by Weyl [21], the Lorentz force in electromagnetic 
theory [22], an absence of the general integrability of the Schr\H{o}dinger equations, 
nonmetric cross-sections in the Yang-Mills theory, the Lagrange function in 
the variational methods, an absence of the identical integrability on equations 
of the mathematical physics and that of identical covariance of the tensor 
equations, and so on. Only if we assume elements of noncovariance, we can obtain 
the closed {\bf inexact} forms. Precisely such closed forms can correspond to the 
physical structures that constitute the physical fields. 

However, the existing field theories are invariant ones because they are provided 
with  
additional conditions, i.e. the conjugation conditions under which the invariance or 
covariance requirements have to be obeyed. Examples of such conditions may be 
the identity relations: 
canonical relations in the Schr\H{o}dinger equations, gauge invariance in electromagnetic 
theory, commutator relations in the Schr\H{o}dinger theory, 
Christoffel's coefficients of connectivity and identity relations by Bianchi 
in the Einstein theory, cotangent bundles in the Yang-Mills theory, the Hamilton 
function in the variational methods, the covariance conditions in the tensor 
methods, the characteristic relations (integrability conditions) in equations 
of mathematical physics, etc. It can be shown that these invariance conditions 
are that of closure for the exterior or dual forms.

As all existing field theories are subjected to the invariance conditions, 
equations of these theories have only invariant solutions, i.e. those that 
can be expressed in terms of the closed or dual to them forms and correspond 
to physical fields. Possible noninvariant solutions, that may be realized 
if the invariance conditions are not imposed, are not accounted for. From this 
standpoint it needs to be recognized two types of equations or methods of 
invariant field theory. A principal distinction of these two types consists 
in the following. One type has only invariant solutions, the other one, if the 
invariance conditions to be omitted, may have noninvariant solutions 
(functionals) that have a physical sense. Equations of the tensor, spinor, 
matrix or variational forms, the group theory methods, the transformation 
theory, theory of symmetries, the bundle theory can be assign to the first 
type. To the second type we may relate the so-called equations of mathematical 
physics, namely, differential and integral equations that describe physical 
processes. As it will be shown later, noninvariant solutions to the second 
type equations play a governing role in description of a mechanism of the 
physical structure creation. At the same time, just equations and methods of 
the first type (equations by Maxwell, Schr\H{o}dinger, Dirac, Einstein, field theory by 
Yang-Mills, the group theory, theory of symmetries and so on) allow one to 
discover a great variety of physical structures, describe physical fields, and 
understand regularities of the physical world.

The field theories that are based on the exact conservation laws allow to describe 
the physical fields. However,  because these theories are invariant ones they 
cannot answer the question about a mechanism of creating physical structures 
which format the physical fields. The answer may be done only by theory where 
the closure conditions of the forms were not introduced. These conditions 
have to be realized by themselves without external forcing. A realization of 
such conditions means a creation of the closed form, and it will correspond to 
the conservation law and point to  appearance of the physical structure. It is 
evident that an answer to a question concerning a mechanism of creation of the 
physical structures may be given only by the evolutionary theory. As it was shown 
above, such a theory is the evolutionary theory which bases on the balance 
conservation laws for the material systems.

\bigskip
\rightline{\bf Appendix 2}
\centerline {\bf Analyses of principles of thermodynamics}

The thermodynamics is based on the first and second principles of thermodynamics, 
which were introduced as postulates [23]. Let us show that the first 
principle of thermodynamics follows from the balance conservation laws for 
energy and linear momentum and is valid for the case when the heat influx is 
the only external action. It appears to be nonidentical relation, and it 
points out that the balance conservation laws are noncommutative and a state of 
thermodynamic system is nonequilibrum one. The second principle of 
thermodynamics with the equality sign is obtained from the first that under 
a realization of the integrability condition, when a role of the integrating factor 
plays the temperature, and it corresponds to a locally equilibrum state of the 
system while the temperature is realized. The second principle of 
thermodynamics with the inequality sign takes into account a presence of some 
actions other than the heat influx.

As it is well known, the first principle of thermodynamics can be presented 
in the form
$$dE+\delta w\,=\,\delta Q $$
where $dE$ is a change of energy of the thermodynamic system, $\delta w$ 
is a work done by the system (this means that $\delta w$ is expressed in terms 
of the system parameters), $\delta Q$ is an amount of heat putted into the 
system (i.e. external action on the system). As the term $\delta w$ has to be 
expressed in terms of the system parameters and it specifies a real (rather then 
virtual) change, then it can be designated by $dw$, and hence, the first 
principle of thermodynamics will take the form
$$dE\,+\,dw\,=\,\delta Q\eqno(1)$$

What is a difference between the first principle of thermodynamics and 
the balance conservation laws? The balance conservation law for thermodynamic 
system can be written as
$$ dE\,=\,\delta Q\,+\,\delta G \eqno(2)$$
where by $\delta G $ we designate energetic actions with the exception 
of heat influx. For thermodynamic system the balance conservation law for 
linear momentum (a change of linear momentum of the system in its 
dependence on the force mechanical action on the system) can be written as
$$dw\,=\,\delta W \eqno(3)$$
Here $\delta W$ signifies a force (mechanical) action on the system (for 
example, external compression of the system, an influence of boundaries 
and so on).

If to sum relations (2) and (3), then one obtains the relation
$$dE\,+\,dw\,=\,\delta Q\,+\,\delta G\,+\,\delta W \eqno(4)$$
which is just the evolutionary relation for the thermodynamic system. 

By comparison the relation (4), that follows from the balance 
conservation laws for energy and linear momentum, with the relation (1), 
one can see that they coincide if the heat influx is the only external 
action on the thermodynamic system ($\delta W\,=\,0$ and $\delta G \,=\,0$).

Thus, the first principle of thermodynamics is obtained from the balance 
conservation laws for energy and linear momentum. This is analogous to 
the evolutionary relation.

The significance of the first principle of thermodynamics, as well as of the 
evolutionary equation, is that it reveals a nature of interactions of two balance 
conservation laws (rather then it only corresponds to the conservation 
law for energy). As $\delta Q$ is not a differential (closed form) then the 
relation (1), which correspond to the first principle of thermodynamics, as 
well as the evolutionary equation appear to be the nonidentical nonintegrable 
relation. This points to the noncommutativity of the balance conservation 
laws (for energy and linear momentum) and to nonequilibrum state of the 
thermodynamic system. As it follows from the analyses of the evolutionary 
relation, a transition to the locally equilibrum state has to correspond to 
the realization of additional condition (i.e. the integrability condition) 
under which this relation turns out to be identical one. Such an identical relation 
is just a relation which corresponds to the second principle of thermodynamics.

Let us consider the case when the work performed by the system is carried out 
through the compression. Then $dw\,=\,p\,dV$ (here $p$ is the pressure and $V$ 
is the volume) and $dE\,+\,dw\,=\,dE\,+\,p\,dV$. As it is known, the form 
$dE\,+p\,dV$ can become a differential if there is the integrating factor 
(a quantity which depends only on the characteristics of the system) $1/\theta\,=\,pV/R$ 
which is named as the thermodynamic temperature $T$ [23]. In this case the 
form $(dE\,+\,p\,dV)/T$ turns out to be a differential (interior) of some quantity 
that referred to as entropy $S$:
$$(dE\,+\,p\,dV)/T\,=\,dS \eqno(5)$$
(although the form $dE\,+\,p\,dV$  consists of differentials, in the general 
case without the integrating factor it is not a differential because of 
that its terms depend on different variables, namely, the first term is 
determined by variables that specifies the internal construction of elements, 
and the second term depends on variables that specify an interaction 
between elements, for example, a pressure). If the integrating factor 
$1/\theta=T$ to be 
realized, that is, the relation (5) proves to be satisfied, then from the 
relation (1), which corresponds to the first principle of thermodynamics, 
it follows
$$dS\,=\,\delta Q/T \eqno(6)$$
This is just the second principle of thermodynamics for reversible processes. 
It takes place when a heat influx is the only action on the system. If besides  
the heat influx the system undergos any mechanical forcing $\delta W$ (for example, 
an influence of boundaries), then according to the relation (4) from the 
relation (5) we obtain
$$dS\,=\,(dE+p\,dV)/T\,=\,(\delta Q+\delta W+\delta G)/T \eqno(7)$$
from which it follows, that
$$dS\, >\,\delta Q/T \eqno (8)$$
which corresponds to the second principle of thermodynamics for irrever\-sible 
processes.

The relations (6), (8) which can be written as
$$dS\,\geq \,\delta Q/T, \eqno(9)$$
express the second principle of thermodynamics. (It is well to bear in mind 
that the differentials in the relations (5), (6), (8), (9)  are not total ones. 
They are satisfied only in presence of the integrating factor, namely, the 
temperature which depends on the system parameters).

Thus, the first principle of thermodynamics is obtained from the balance conservation 
laws for energy and linear momentum, and the second principle of thermodynamics 
does from the first one. The second principle of thermodynamics with the equality 
sign follows from the first principle under the fulfillment of the condition 
of integrability, i.e. a realization of the integrating factor (temperature). 
(This transition corresponds to that from the nonequilibrum state to the locally 
equilibrum state. Phase transitions, the origin of fluctuations, etc are 
examples of such transitions). And the second principle of thermodynamics with the 
inequality sign takes into account a presence in the real processes other 
actions besides the heat influx.

In the case examined above a differential of entropy (rather then entropy itself) 
becomes the closed form. $\{$In this case entropy  manifests itself as the thermodynamic 
potential, namely, the function of state. To the pseudostructure there corresponds 
the state equation which determines the temperature dependence on the 
thermodynamic variables$\}$. For entropy to be the closed form itself, a one more 
condition has to be realized. Such a condition may be a realization of the 
integrating direction, an example of that is the sound speed:
$a^2\,=\,\partial p/\partial \rho\,=\,\gamma\,p/\rho$. In this case it is valid 
the equality $ds\,=\,d(p/\rho ^{\lambda })\,=\,0$ from which it follows that 
entropy $s\,=\,p/\rho ^{\lambda }\,=\,const$ is the closed form (of zero degree). $\{$But 
it does not mean that a state of the gaseous system is identically isoentropic 
one. Entropy is constant only along the integrating direction (for example, on 
the adiabatic curve or on the front of sound wave), whereas in the direction 
being normal to the integrating direction the normal derivative of entropy 
has a break$\}$.

It worth underline that both temperature and the sound speed are not continuous 
thermodynamic variables. They are variables which are realized in the 
thermodynamic processes if the thermodynamic system has any degrees of 
freedom. One may see an analogy between temperature and the sound speed: 
temperature is the integrating factor and the sound speed is the integrating 
direction. $\{$Notice that in actual processes a total state of the 
thermodynamic system is nonequilibrum one and a commutator of the form 
$dE\,+\,pdV$ is nonzero. A quantity that is described by commutator and 
acts as internal force may grow. Prigogine [19] defined this as the "production 
of the excess entropy". Just this increase of the internal force is perceived 
as the growth of entropy in the irreversible processes$\}$.

The closed static system, if left to its own devices, may tend to a state 
of total thermodynamic equilibrum. This corresponds to tending of the system 
functional to its asymptotic maximum. In the dynamical system the tending 
of the system to a state of total thermodynamic equilibrum may be violated by 
dynamical processes and transitions to a state of local equilibrum.

\bigskip
\rightline{\bf Appendix 3}

{\bf Influence of noncommutativity of the balance conservation laws
on a development of instability}.

A noncommutativity of the balance conservation laws, that leads to the 
emergence of internal forces and an appearance of the nonequilibrum, is a 
cause of development of instability in the material systems. Hence, to study 
a cause of development of instability, it is necessary to examine the 
evolutionary relation obtained from the balance conservation laws and analyze 
a commutator of the nonintegrable form entered into this relation.

For example, we take the simplest gas dynamical system, namely, a flow 
of the ideal (nonviscous, heat nonconductive) gas [24]. Suppose, that gas 
is the thermodynamic system in the local equilibrum (whereas the gas 
dynamical system may be in the nonequilibrum state). This means that the following relation 
is satisfied [23]
$$Tds\,=\,de\,+\,pdV \eqno(1)$$
here $T$, $p$ and $V$ are the temperature, the pressure, and the gas volume respectively, 
$s$, $e$ are entropy and the internal energy per unit volume of the gas.

Let us introduce two frames of reference: the inertial one, which is not connected 
with the material system, and the accompanying system that is connected 
with a manifold formatted by trajectories of elements of the material system. 
(As examples these may be both Euler's and Lagrange's systems of coordinates). 

In the inertial system of coordinates the Euler equations are the balance 
conservation laws for energy, linear momentum and mass of the ideal gas [16].
The balance conservation law for energy of the ideal gas can be written as
$${{Dh}\over {Dt}}- {1\over {\rho }}{{Dp}\over {Dt}}\,=\,0$$
where $D/Dt$ is the total derivative with respect to time (if to designate 
the spatial coordinates by $x_i$ and velocity components by $u_i$, then 
$D/Dt\,=\,\partial /\partial t+u_i\partial /\partial x_i)$,  $h$ is the entalpy  
of the gas. Expressing the entalpy in terms of the internal energy $e$ with 
the help of the formula $h\,=\,e\,+\,p/\rho $ and using the relation (1) we can 
reduce the equation of balance conservation law for energy to the form 
$${{Ds}\over {Dt}}\,=\,0 \eqno(2)$$
And respectively, the equation of balance conservation law for momentum can be 
presented as [16, 25]
$$\hbox {grad} \,s\,=\,(\hbox {grad} \,h_0\,-\,{\bf U}\times \hbox {rot} {\bf U}\,+\,{\bf U}\times {\bf F}\,+\, 
\partial {\bf U}/\partial t)/T \eqno(3)$$
where ${\bf U}$ is the velocity of the gas particle, $h_0=({\bf U \cdot U})/2+h$, 
${\bf F}$ is the mass force. 

As the total derivative with respect to time is that along the trajectory, 
then in the accompanying frame of reference the equations (2), (3) take the form:
$${{\partial s}\over {\partial \xi ^1}}\,=\,0 \eqno (4)$$
$${{\partial s}\over {\partial \xi ^{\nu}}}\,=\,A_{\nu },\quad \nu=2, ... \eqno(5)$$
where $\xi ^1$ is the coordinate along the trajectory, 
$\partial s/\partial \xi ^{\nu }$ 
is the left-hand side of the equation (3), and $A_{\nu }$ is obtained from the 
right-hand side of the relation (3). $\{$In the common case when gas is 
nonideal the equation (2) can be written in the form 
$${{\partial s}\over {\partial \xi ^1}} \,=\,A_1 \eqno (6)$$
where $A_1$ is the expression which depends on the energetic actions. In the case 
of ideal gas $A_1\,=\,0$ and the equation (6) transforms into (4). In the case 
of viscous heat-conductive gas, described by a set of the Navier-Stokes equations, 
in the inertial coordinate system can be written in the form [16]
$$A_1\,=\,{1\over {\rho }}{{\partial }\over {\partial x_i}}
\left (-{{q_i}\over T}\right )\,-\,{{q_i}\over {\rho T}}\,{{\partial T}\over {\partial x_i}}
\,+{{\tau _{ki}}\over {\rho }}\,{{\partial u_i}\over {\partial x_k}} \eqno(7)$$
Here $q_i$ is the heat flux, $\tau _{ki}$ is the viscous stress tensor. In the case 
of reacting gas there are added additional terms connected with the chemical 
nonequilibrum.

The equations (4) and (5) can be convoluted into the equation
$$ds\,=\,\omega \eqno(8)$$
where $\omega\,=\,A_{\mu} d\xi ^{\mu}$ is the differential form of the first 
degree (here $\mu =1,\,\nu $).

The relation (8) is the evolutionary relation for the gas dynamical system 
(in the case of the local thermodynamic equilibrum). Here $\psi\,=\,s$. 
$\{$It worth notice that in the evolutionary relation for the thermodynamic 
system it is examined a dependence of entropy on the thermodynamic variables 
(see the relation (1)), but in the evolution relation for the gas dynamical system
the entropy dependence on the space-time variables is considered$\}$. If the relation 
(8) be identic one (if the form $\omega $ be the closed form, i.e. a differential), 
then one can obtain  differential of entropy $s$ and find entropy as a function 
of space-time coordinates. Just this entropy will be the gas dynamical function 
of state, a presence of which will point to that a state of the gas dynamical 
system is locally equilibrum. And if the relation (8) will not be identical, then 
the differential of entropy cannot be defined a differential of entropy $s$. 
This will point to an absence of the gas dynamical function of the state and nonequilibrum 
state of the system. Such nonequilibrum is a cause of a development of instability.

Because of that nonequilibrum is produced by internal forces which are described 
by a commutator of the form $\omega $, then it becomes evident that a cause 
of the gas dynamical instability is something that contributes to a commutatior 
of the form $\omega $. Without an accounting for terms which are connected with 
a deformation of the manifold formatted by trajectories, the commutator can be 
written as
$$K_{1\nu }\,=\,{{\partial A_{\nu }}\over {\partial \xi ^1}}\,-\,{{\partial A_1}\over 
{\partial \xi ^{\nu }}} \eqno(9)$$

From the analyses of the expression $A_{\nu }$ and with taking into account 
that $A_1\,=\,0$, one can see, that into commutator there contribute terms 
which are related to the multiple connectivity of the flow domain (the second term of 
the expression for $A_{\nu }$), nonpotentiality of the external forces 
(the third term) and a nonstationarity of the flow (the forth term). $\{$In 
the general case the terms connected with the transport phenomena and the 
physical and chemical processes contribute to the commutator (9)$\}$. These 
factors lead to the origin of internal forces, to nonequilbrum state 
and to the development of instability of various kinds. And yet for every 
kind of instability one can find an appropriate term in the commutator of the 
nonintegrable form, which is responsible for this kind of instability. 
$\{$Under this process it is also necessary to consider the evolutionary relations 
which the balance conservation laws for angular momentum and mass correspond 
to$\}$. Thus, there is an unambiguous connection between a kind of instability 
and the terms that contribute into the commutator of the nonintegrable form 
in the evolutionary relation. 

As it was shown, under the realization of the additional degrees of freedom, 
a development of instability may lead to origin of the physical 
structures when the internal forces transform into the potential ones. 
Such gas dynamical structures are shocks, shock waves, turbulent pulsations 
and so on. Additional degrees of freedom are realized as the condition of 
the degenerate transform, namely, vanishing of determinants, Jacobians of 
transforms, etc. For example, such conditions are justified either on 
characteristics (determinant of coefficients at the normal derivatives 
vanishes), or at the singular points (Jacobian is equal to zero) on the 
envelop of the characteristics of the Euler equations, or at the singular 
points of the Navier-Stokes equations (when the viscous gas is considered).

Let as analyze what kinds of instability and what gas dynamical structures 
may originate under given forcing.

1). {\it Shock, break of diaphragm and others}. The instability originates because 
of nonstationarity. The last term in the equation (3) gives a contribution 
into the commutator. In the case of ideal gas, whose flow is described by 
equations of the hyperbolic type, a transition to the locally equilibrum 
state is possible on the characteristics and their envelops. The corresponding 
structures are weak shocks and shock waves.

2).{\it Flow around bodies by ideal (unviscous, thermal unconductive) gas. 
Action of the nonpotential body forces}. The instability develops because of 
the multiple connectivity of the flow domain and the nonpotentiality of the 
body forces. A contribution into the commutator has come from the second and third 
terms of the right-hand part of the equation (3). As the gas is ideal one and 
$\partial s/\partial \xi ^1=A_1=0$, that is there is no contribution into 
every fluid particle, then the instability of convective type develops. 
For $U>a$ ($U$ is the velocity of the gas particle, $a$ is the speed of sound) 
a set of equations of the balance conservation laws belongs to the hyperbolic type and hence a 
transition to the locally equilibrum state is possible on the characteristics 
and the envelops of characteristics as well, and weak shocks and shock waves are the 
structures of the system. If $U<a$ when the equations are of elliptic type, such a 
transition is possible only at singular points. The structures, originated because of 
the convection, are of the vortex type. Under long forcing the large-scale structures 
can be produced.

3. {\it Boundary layer}. The instability originates because of the multiple 
connectivity of the domain and the transport phenomena (an influence of 
viscosity and thermal conductivity). Contributions into the commutator produce the second term 
in the right-hand part of the equation (3) and the second and third terms 
in the expression (7). A transition to the locally equilibrum state is allowed 
at the singular points. As in this case $\partial s/\partial \xi^1=A_1\neq 0$,
that is, forcing acts on every gas particle singly, then a development of 
instability and transitions to the locally equilibrum state are allowed 
only in single fluid particle. Hence, the structures originated behave as pulsations. 
These are the turbulent pulsations.

In worth notice that separating of some formation from the local domain 
is accompanied by a production of the discontinuous surfaces (the contact discontinuities). 
Unlike shocks and shock waves these discontinuities do not propagate relative 
to the material system.

\bigskip
\rightline{\bf Appendix 4}
\centerline{\bf Electromagnetic field}

We will show how it may be obtained the evolutionary relation for the system 
which generates the electromagnetic field and what are
specific features of this relation. 

As it is known [11], the electromagnetic field can be presented by some 
2-form and its dual one. The Maxwell equations appear to be reduced to 
that the both forms are closed forms. Let us analyze when the closure 
conditions are satisfied.

If to utilize the Lorentz force ${\bf F\,= \,\rho (E + [U\times H]}/c)$, 
then a local variation of energy and linear momentum of the charged 
substance may be written respectively as [17]: $\rho ({\bf U\cdot E})$, 
$\rho ({\bf E+[U\times H]}/c)$. Here ${\bf E\,,H}$ are respectively the 
electric and magnetic strengths of the field, $\rho$ is the charge 
density, ${\bf U}$ is the velocity of the charged substance. These 
variations of energy and linear momentum are caused by energetic and 
force actions and are equal to values of these actions. If to denote 
these actions by $Q^e$, ${\bf Q}^i$, then the balance conservation laws 
can be written as follows:
$$\rho \,({\bf U\cdot E})\,=\,Q^e$$
$$\rho \,({\bf E\,+\,[U\times H]}/c)\,=\, {\bf Q}^i \eqno (1)$$

After elimination of the characteristics of the material system (charged 
substance) $\rho$ and ${\bf U}$ by application of the Maxwell-Lorentz equations 
[17], the left-hand sides of the equations (1) can be expressed only through 
the strengths of the electromagnetic field  and then one can write the 
equations (1) in the form:
$$c\,\hbox{div} {\bf S}\,=\,-{{\partial}\over {\partial t}}\,I\,+\,Q^e\eqno(2)$$
$${1\over c}\,{{\partial }\over {\partial t}}\,{\bf S}\,=
\,{\bf G}\,+\,{\bf Q^i}\eqno(3)$$
where ${\bf S=[E\times H]}$ is the Pointing vector, $I=(E^2+H^2)/c$, 
${\bf G}={\bf E}\,\hbox {div}{\bf E}+\hbox{grad}({\bf E\cdot E})-
({\bf E}\cdot \hbox {grad}){\bf E}+\hbox {grad}({\bf H\cdot H})-({\bf H}\cdot\hbox{grad}){\bf H}$. 

The equation (2) is widely utilized while description of the electromagnetic 
field and a calculation of energy and the Pointing vector. And the equation (3) 
does not commonly be taken into account. Actually, the Pointing vector 
${\bf S}$ has to obey to two equations, which can be convoluted into the 
{\bf relation}
$$d\,\psi\,=\,\omega ^2\eqno (4)$$
where 2-form $d\,\psi\,$ corresponds to the vector $\bf S$ and coefficients of the form $\omega ^2$ 
(the upper subscript shows a degree 
of the form) are the right-hand parts of the equations (2), (3). It is just 
the evolutionary relation for the system of charged particles that generates 
the electromagnetic field.

From the equations (2), (3) or from the evolutionary relation one can find 
the Pointing vector as some preservable measured physical quantity only if 
these equations are conjugated ones, that is, if a commutator formed by mixed 
derivatives (it is just a commutator of the form $\omega ^2$) is equal to zero. 
And if the commutator is nonzero, then the right-hand side is not a 
differential and the Pointing vector is some functional. Under what conditions 
can the Pointing vector be formatted as a measurable quantity?

Let us choose the local coordinates $l_k$ in such a way that one direction 
$l_1$ coincides with a direction of the vector ${\bf S}$. As this chosen direction coincides with a 
direction of the vector ${\bf S=[E\times H]}$ and hence is normal to the vectors 
${\bf E}$ and ${\bf H}$, 
then one obtains that $\hbox{div} {\bf S}\,=\,\partial s/\partial l_1$, where $S$ is 
the module of ${\bf S}$. And in addition, 
the projection of the vector ${\bf G}$ on the chosen direction turns out to be 
equal to $\partial I/\partial l_1$.
As the result, after separating from the vector equation (3) its projection 
on the chosen direction, the equations (2), (3) can be written as 
$${{\partial S}\over {\partial l_1}}\,=\,-{1\over c}{{\partial I}\over {\partial t}}\,-\,
{1\over c}Q^e \eqno(5)$$
$${{\partial S}\over {\partial t}}\,=\,c\,{{\partial I}\over {\partial l_1}}\,-\,c{\bf Q}'^i\eqno(6)$$
$$0\,=\,-{\bf G}''\,-\,c{\bf Q}''^i$$
Here the prime relates to the direction $l_1$, double primes do to the other 
directions. Under the condition $\partial l_1/\partial t\,=\,c$ from the 
equations (5), (6) it is possible to obtain a relation in the differential 
forms
$${{\partial S}\over {\partial l_1}}\,dl_1\,+\,{{\partial S}\over {\partial t}}\,dt\,=\,
-\left( {{\partial I}\over {\partial l_1}}\,dl_1\,+\,{{\partial I}\over {\partial t}}\,dt\right )\,-\,
(Q^i\,dt\,+\,{\bf Q}'^e\,dl_1)\eqno(7)$$
As the second brace in the right-hand side is not a differential (the energetic and force 
actions have distinguished nature and cannot be conjugated), then one can obtain 
a closed form only if the right-hand side vanishes:
$$\left ({{\partial I}\over {\partial l_1}}\,dl_1\,+\,{{\partial I}\over {\partial t}}\,dt\right )
\,-\, (O^i\,dt\,+\,{\bf Q}'^e\,dl_1)\,=\,0\eqno(8)$$
that is
$${{\partial I}\over {\partial t}}\,=\,{\bf Q}'^e,\quad {{\partial I}\over {\partial l_1}}\,=\,Q^i\eqno (9)$$
In this case $dS\,=\,0$ and the module of the Pointing vector $S$ proves to be 
the closed form, i.e. a measurable quantity. And the integrating direction 
(pseudostructure) will be
$$-\,{{\partial S/\partial t}\over {\partial S/\partial l_1}}\,=\,{{dl_1}\over {dt}}\,=\,c\eqno(10)$$
Thus, the constant $c$, that was introduced into the Maxwell equations, is defined 
as the integrating direction. 
From the expressions (9) it is evident, that in this case the local energetic and force 
actions on the material system (charged substance) appear to be transformed into the quantities 
of the electromagnetic field, namely, energy and linear momentum of the electromagnetic wave that 
propagates with the light speed $c$, which value is defined by the condition 
(10). One can see, that the constant $c$ in the Maxwell equation is the speed 
of the electromagnetic wave and it is defined as the integrating direction. 

\bigskip
\rightline{\bf Appendix 5}

\centerline {\bf On interactions}

As it was shown above, a type of the physical structures (and, accordingly, the physical 
fields) generated by the evolutionary relation depends on a degree of the 
exterior forms $p$ and $k$ (here $p$ is a degree of the nonintegrable form 
of the evolutionary relation which is connected with a number of the interacting 
balance conservation laws, and $k$ is a degree of the closed form generated 
by the evolutionary relation) and on a space dimension $n$. By introducing 
the classification by numbers $p$, $k$, $n$ one may understand the internal connection of 
different physical fields and see a connection between interactions. It is reflected in 
the table presented below.

In the table names of the particles created are given. Numbers placed near particle 
names correspond to the space dimension. In the curly brackets the 
sources of interactions are presented. In the lower row we point out the 
massive particles created by interactions (the exact forms of zero degree obtained by 
sequent integrating of the forms of degree $p$ correspond to these particles). 
From the table one can see a correspondence between the degree $k$ of the closed forms 
being realized and a type of interactions. Thus, $k=0$ corresponds to the strong 
interaction, $k=1$ does to the weak one, $k=2$ corresponds to the 
electromagnetic interaction, and $k=3$ does to the gravitational interaction. As a result, 
we obtain that a type of interaction 
and a type of the created particles is defined by the degree ($k$) of the closed 
forms realized, and properties of the particles are governed by the degree ($p$) 
of the evolutionary equation, namely, by a number of the interacting balance conservation 
laws, and the space dimension. The last property is connected with that the 
closed forms of equal degrees $k$, but obtained from the evolutionary 
relations acting in spaces of different dimensions $n$ (the values of $p$ are 
different), are distinguished as they are defined on the pseudostructures 
of different dimensions (the dimension of the pseudostructure $(n+1-k)$ depends on the 
dimension of the initial space $n$). For this reason the realized physical structures with equal degrees $k$ of the closed forms are 
distinguishable. A connection between a type of interactions and the conservation laws may 
be seen by a comparison of the first and last columns. In the last column the interacting 
balance conservation laws are pointed out. The arrows show that the 
conservation laws presented in given cell are added to that from lower cells.
Notice that for $k=0$ in the space of zero dimension 
there is no the momentum. 
This manifests it self beginning with the space of the dimension 1 (in the forms of degree $k=0$ energy 
and momentum are formatted independently). For $k=0$ energy and momentum are nonconjugate. 
The forms dual to them are respectively time and coordinates (time and coordinates have 
different nature because time is dual to energy and coordinates are dual to 
linear momentum). Energy and time as well as linear momentum and coordinate do not 
commutate within the framework of the same form because they relate 
to different forms: exterior and dual ones. (The commutative relations 
$\hat q\hat p -\hat p \hat q=\imath \hbar $ reflect this fact. The left-hand side of 
the commutative relations is analogous of the commutator value of the nonintegrable 
form of zero degree, and the right-hand side is equal to its value at the 
point in time of the realization of the closed zero degree form, the imaginary unit 
points to the transverse direction with respect to the pseudostructure). The closed exterior form 
of degree $k=1$ corresponds to a conjugation of energy and momentum (for $k=1$ energy and 
momentum prove to be the simultaneously measurable quantities). 

$\{$In the table a single cycle of formation of the physical structures is 
presented. It contains four levels to every of those it corresponds a proper 
value $p$ ($p=0,1,2,3$) and the space dimension $n$. The structures formatted 
in the preceding cycle serves as a source of interactions for the first 
level of a new cycle. The sequential cycles reflect properties of the sequentially 
embedded material systems. And some given level has specific properties 
which are inherent in the same levels of the other cycles. For example, it may 
be seen by comparison of the cycle described and the cycle where conductors, 
semiconductors, dielectrics and neutral elements are sequentially correspond 
to the exact forms. Properties of elements of the third level, namely, neutrons 
in one cycle and dielectrics in the other coincide with those of the so-called 
"magnetic monopole" [26, 27]$\}$.

\vfill\eject
\vbox {\baselineskip=12pt
\rightline{\bf TABLE}
%\bigskip
\halign{\qquad\bf#\hfil&\quad\hfil#\hfil&\quad\bf#\hfil&\quad\bf#\hfil&\quad\bf#\hfil
&\quad\bf#\hfil&\quad#\hfil\cr
interaction&$\backslash$\rm n&\rm i+0&\rm i+1&\rm i+2&\rm i+3&\bf balance\cr
&$\rm k\backslash p$&0&1&2&3&\bf conserv.\cr
&&&&&&\bf laws\cr
&&&&&&\cr
\noalign{\hrule}
\noalign{\hrule}
\noalign{\hrule}
&&&&&&\cr
gravitation&&&&&graviton&\cr
&\bf3&&&&$ \Uparrow \{$\rm electron&mass\cr
&&&&&\rm +proton&\cr
&&&&&\rm +neutron &\cr
&&&&&\rm +photon &\cr
&&&&&\rm (neutrino &\cr
&&&&&\rm +quant)?&\cr
\noalign{\hrule}
&&&&&&\cr
electro-&&&&photon 2&photon 3&angular\cr
magnetic&\bf2&&&$\Uparrow$&&momentum\cr
&&&&\rm$\{$electron+&&\cr
&&&&\rm proton+&&\cr
&&&&\rm neutrino$\}$&&$\uparrow$\cr
&&&&\rm $\{$quant$\}$?&&\cr
&&&&&&\cr
\noalign{\hrule}\cr
weak&\bf1&&neutrino 1&neutrino 2&neutrino 3&linear \cr
&&&$\Uparrow$&&&momentum\cr
&&&$\{$\rm electron&&&\cr
&&&\rm +quant $\}$&&&$\uparrow$ \cr
&&&&&&\cr
\noalign{\hrule}\cr
&&&&&&$p=0$\cr
strong&\bf0&quant 0&quant 1&quant 2&quant 3&energy+\cr
&&$\Uparrow$&\rm muons?&&&time\cr
&&\rm (quarks?)&\rm gluons?&&&or $p>0$\cr
&&&&&&momentum\cr
&&&&&&+coordin.\cr
&&&&&&\cr
\noalign{\hrule}
\noalign{\hrule}
\noalign{\hrule}
&&&&&&\cr
particles&&electron&proton&neutron&deuteron ?&\cr
\rm material&\rm exact&&&&\cr
\rm (nuclons?)&\rm forms&&&&&\cr}}

\bigskip 
\rightline{\bf Appendix 6}
\centerline{\bf A formation of the metric space}

As it was shown above, a noncommutativity of the balance conservation laws and 
a transition from those to the exact conservation laws explain a mechanism of 
origin of the physical structures. And as the origin of the physical 
structures is connected with a formation of the pseudostructures, then by analyzing 
a mechanism of formation of the physical structures one can understand 
a mechanism of formation of the pseudometric and metric spaces. $\{$Recall, that 
the inexact closed forms correspond to the physical structures and pseudostructures.
and the exact forms correspond to elements of the material system and to the 
metric spaces$\}$.
It is useful to note some properties of the manifolds.

Assume, that on the manifold one can choose any system of coordinates 
with the basis ${\bf e}_{\mu }$ and to set up the metric forms of manifold [21]:  
$({\bf e}_{\mu }\,{\bf e}_{\nu })$, $({\bf e}_{\mu }\,dx^{\mu })$, $(d\,{\bf e}_{\mu })$. 
The metric forms and their commutators define metric and differential 
characteristics of manifold. If the metric forms are closed 
(commutators are equal to zero), then the metric $g_{\mu \nu }=({\bf e}_{\mu}{\bf e}_{\nu })$ 
is defined, and the results of 
displacement over manifold the point $d{\bf M}=
({\bf e}_{\mu}dx^{\mu })$ and a unit vector $d{\bf A}=(d{\bf e}_{\mu })$ 
are independent of the path of displacement (the path of integration). 
The closed metric forms define the structure of the manifold, and the commutators 
of the metric forms determine such characteristics of the manifold as bend, 
torsion, curvature (if the commutators are nonzero, then the results of 
displacement over manifold of a point or the unit vector depend on the path 
of displacement). An example of manifold with the closed metric forms is the 
differentiable manifold whose metric and differential characteristics 
prove to be consistent [13].

A role of the differential characteristics of the manifold may play 
the connectivities [21] $\Gamma _{\mu \nu }^{\rho }$, 
$(\Gamma _{\mu \nu }^{\rho }-\Gamma _{\nu \mu }^{\rho })$, 
$R_{\nu \mu \sigma }^{\mu }$, {which are components of a commutator of the metric forms. 

As it is known [20], these commutators for the Euclidean manifold are equal to zero.
In the case of the Riemann manifold the commutator of the metric form of degree 
two is nonzero: $R_{\nu \rho \sigma }^{\mu }\neq 0$.

If the exterior differential forms  are defined on manifolds whose metric forms are 
nonclosed, then, as it was pointed out, the commutators of the metric forms will enter 
into the commutators of the exterior differential forms. In particular, 
components of the commutator of the external form of degree one 
$\theta\,=\,a_{\alpha }\,dx^{\alpha }$ can be written 
in terms of connectivities as follows:
$$K_{\alpha \beta }\,=\,\left ({{\partial a_{\beta }}\over {\partial x^{\alpha }}}
\,-\,{{\partial a_{\alpha }}\over {\partial x^{\beta }}}\right )\,+\,
(\Gamma _{\beta \alpha }^{\sigma }-\Gamma _{\alpha \beta }^{\sigma })\,a_{\sigma }\eqno(1)$$

It is evident, that the commutator of the exterior form consists of two terms. 
The first one (the first parentheses) depends on coefficients of the exterior form, 
and the second term does on the differential characteristics of the manifold. 
It is a typical feature for the commutator of the exterior forms of other degrees as well.
Because of this feature the topologic properties of commutators of the exterior forms 
are manifested: they can realize a mutual relation between the exterior form and 
the basis, namely, the metric form of manifold. The other specific feature is 
that terms in the commutator have the different nature, 
i.e. one term depends on coefficients of the exterior form, and the other depends on 
the basis. Such terms cannot be equal identically, and hence they cannot 
make the commutator to be zero. This means that the exterior forms, defined 
on manifolds with nonclosed metric forms, turns out to be nonclosed (such forms 
may be called nonintegrable ones in contrast to the nonclosed forms defined 
on the differentiable manifolds). The above mentioned properties of commutators of 
the exterior forms enable one to understand a mechanism of formation of the 
metric spaces. Note, that the material system is the generator  of the 
formatting metric space, and an accompanying manifold is the base. 

When derivation the evolutionary relation there were used two spatial objects: 
the accompanying manifold (connected with the material system) that has no metric 
structure for real processes and the inertial space (not connected with 
the material system) which is the metric space. $\{$Note, that the metric space formatted 
becomes the inertial space of one more degree$\}$.

Assume that the initial inertial space has the dimension $n=3$. The material system 
in such a space is subjected to the balance conservation laws, which equations   
in the accompanying frame of reference turn out to be convoluted to the evolutionary relation 
with $p=n=3$:
$$d\,\psi\,\cong \,\omega ^3,\quad d\,\omega ^3\,\neq \,0 \eqno(2)$$

The form $\omega ^3$ is defined on the accompanying manifold, and therefore 
this form is nonintegrable, that is, its commutator is nonzero (a degree 
of the form $\psi $ equals 2). $\{$In cosmology and the gravitation theory 
the equations of ideal fluid are sometimes used [15, 17, 28]. In essence,
these equations are the balance conservation laws. However, these equations 
commonly used in the covariant form [28], that is, the covarince condition is 
imposed. To study a process of formation of the pseudometric and metric 
spaces it is necessary to employ equations that are not subjected to the 
conditions of invariance or covariance$\}$.

A realization of the pseudostructure (an element of the pseudometric 
space)
and an origin of the physical structure, which  the closed metric and exterior 
forms correspond to, is the transition from the nonintegrable form $\omega ^3$ 
to the closed form $\omega '^3$ (this is connected with the degenerate transform). 
And it is required the following relations have to be satisfied:

$$d_{\pi }\,\omega '^3\,=\,0 $$
$$d_{\pi }\,^*\omega \,'^3\,=\,0\eqno(3)$$
In the present case a degree of the closed form is $k=p=3$, and a dimension 
of the pseudostructure is $m=n+1-k=p+1-k=1$. On the pseudostructure from the 
evolutionary relation (2) it follows the relation
$$d_{\pi }\,\psi \,=\,\omega '^3 \eqno(4)$$
which is identical one because the closed form $\omega '^3$ may be expressed 
through the interior differential. From this relation it can be defined the form $d_{\pi }\psi $ 
which specifies a state of the system and may be referred to as the 
structural form. (In the case under consideration this is the form of degree 3). 
It corresponds to the conservation law, because a differential of this form 
(interior on the pseudostructure) is equal to zero.
 
A realization of the physical structure (connected with an origin of the 
physical structure and that the conservation law is fulfilled) is one of the exhibitions 
of a mechanism of formatting the metric spaces. It worth to underline that the 
pseudostructure is realized with respect to the inertial frame of reference. 
(The degenerate transform corresponds to transition from the frame of reference 
connected with the accompanying manifold to the inertial coordinate system).

With the aim to be more clear we shall put the tensor expressions into correspondence 
to the exterior forms. We may put a tensor with $p$ lower (covariant) 
subscripts into correspondence to the external form of degree $p$ defined on 
the differentiable manifold. As it is known, a differential of the form degree 
$p$ on the differentiable manifold is the form of degree $p+1$. We may put a tensor 
with $p+1$ lower subscripts into correspondence to the differential of the form 
of degree $p$. By analogy with this we put the tensor expression 
$K_{\alpha ...}$ into correspondence to a differential or to a commutator of the 
nonintegrable form. With this notation a commutator of the form $\omega ^3$ 
can be written as $K_{\alpha \beta \gamma \chi }$, where three first subscripts 
corresponds to a degree of the form, and the fourth one appears while differentiating the form 
(from this point and further we shall use the Greek subscripts for the 
accompanying frame of reference and Latin ones for the inertial that). 
A commutator of the basic metric form which can be denoted by $R_{\alpha \beta \gamma \chi }$ 
enters into a commutator of the nonintegrable form. We may put the covariant tensors of 
rank 3, $S_{jkl}$ and $T_{jkl}$, (its divergence is equal to zero as they 
corresponds to the closed forms) into correspondence to the closed forms 
$d_{\pi }\psi$ and $\omega '^3$ 
(that are formatted with relevance to the inertial frame of reference). And 
to the pseudostructure we may put into correspondence  
the 1-covariant pseudotensor $T^i$ (it corresponds to the closed metric 
form, i.e. pseudostructure), which is dual to the tensor $T_{jkl}$[15]:
$T^i\,=\,^*T_{ijk}\,=\, {1\over 6}\varepsilon ^{ijkl}\,T_{jkl}$ (here 
$\varepsilon ^{ijkl}\,=\,e_ie_je_ke_l\varepsilon _{ijkl}$, where $\varepsilon _{ijkl}$ 
is the completely antisymmetric unit pseudotensor) can be put into correspondence to the 
pseudostructure. Similarly, by $S^i\,=\,^*S_{jkl}$ denote the tensor dual to 
$S_{jkl}$. Now we introduce the tensor expressions:
$$
{\bf S}_{jkl}^i\,=\,\cases{S_{jkl} \cr S^i },\quad {\bf T}_{jkl}^i\,=\,
\cases{T_{jkl}\cr T^i}
\eqno(5)$$
$\{$These tensor expressions are not tensors with covariant and contravariant 
subscripts because, firstly, they combine tensors and pseudotensors, and secondly, 
in these expressions one cannot rise up and lower subscripts as the metric is not 
defined as yet$\}$. The tensor expression ${\bf S}_{jkl}^i$ is a representation 
of the physical structure (it corresponds to the structure form on 
pseudostructure), and the tensor expression ${\bf T}_{jkl}^i$ is a representation 
of the forms $\omega '^3$ and $^*\omega '^3$ (they correspond to the external 
actions processed by the system).

With taking into account the relations (3), the relation (4) can be written in terms of 
the tensor expressions as
$${\bf S}_{jkl}^i\,=\,{\bf T}_{jkl}^i\eqno(6)$$

The relation (6) shows that the physical structure (including the 
pseudostructures) are produced at the expense of external actions processed 
by the system.

What is the further mechanism of formation of the metric space?

While origin of the physical structure, a quantity, which is described 
by a commutator of the nonintegrable form $\omega ^3$ and acts as 
internal force, transforms into potential force, which acts in the direction 
transverse to the pseudostructure. (If a differential of the form $\omega ^3$ 
be zero, that is, the commutators $R_{\alpha \beta \gamma \chi }$ and 
$K_{\alpha \beta \gamma \chi }$ be equal to zero, 
then the potential force will be equal to zero). This potential force becomes a new source 
of noneqilibrum (even without the extra external actions) and may lead to 
a further formation of the pseudostructures.

As the relation (4) is the identical one, then it can be integrated. Because the 
form $\omega '^3$ is closed, it is the interior (on the pseudostructure) 
differential of the form one degree less
$$\omega '^3\,=\,d_{\pi }\omega ^2\eqno(7)$$
From the relations (4), (7) it follows the relation (below, for the sake of convenience, 
we shall indicate explicitly a degree of the form $\psi $)
$$d_{\pi }\,\psi ^2\,=\,d_{\pi }\,\omega ^2$$
which can be integrated (within the accuracy up to the lower degree forms):
$$\psi ^2\,=\,\omega ^2\eqno(8)$$
This is an integration of the nonidentical evolutionary relation (2) over a 
single dimensionality which has been formatted.

From the relation (7) one can see that a differential of the form $\omega ^2$ 
is nonzero. The form $\omega ^2$ (of degree $p-1=2$) proves to be nonintegrable 
form (its commutator is nonzero) on the manifold directions remained after 
integration. To the commutator of the form $\omega ^2$ it can be put into 
correspondence the tensor expression $K_{\beta \gamma \chi }^{\alpha }$ (three lower subscripts is a degree 
of the exterior form plus 1, and a single top subscript is the pseudometric 
dimension formatted). In this case the basic commutator can be written in the form 
$R_{\beta \gamma \chi }^{\alpha }$. (If the Bianchi identity [21] is satisfied, 
then from this tensor it can be obtained the Riemann-Christoffel tensor 
$G_{jkl}^i$ which corresponds to the Riemann manifold. 
However it takes place only after the pseudoriemann and Riemann manifolds 
be completely formatted).

Here it appears some special feature. On the one hand, the form $\omega ^2$ obtained  
turns out to be nonintegrable one, and therefore, it cannot be expressed in terms 
of differential. And on the other hand, the form $\psi ^2$ in the left-hand side 
of the relation (8), for to become the structural form, has to become the closed 
form, namely, the differential:
$$\psi ^2\,=\,d\,\psi ^1\eqno(9)$$
By comparison of the relations (8) and (9), we get
$$d\,\psi ^1\,\cong \,\omega ^2\eqno(10)$$
which cannot be identity as the form $\omega ^2$ is not expressed through 
differential.

The nonidentical relation (10) is the relation of the type 
similar to the initial relation (2), however it is the form 
of one less degree. We may repeat the analysis like for 
the relation (2) and get the pseudostructure of one more dimension. 
By sequent integrating of the nonidentical relations we may obtain 
the pseudometric space. The closed exterior forms of degrees 
$p,\,p-1,\,...,\,0$, which are inexact, correspond to this space. 
A transition to the exact form of zero degree will correspond to 
a transition to the metric space.

With application of the tensor expressions these transitions 
can be schematically written in the following form:

%\vfill\eject
$$d\psi \cong \omega ^3, \quad d\omega ^3 \neq 0\quad (K_{\alpha \beta \gamma \chi}\neq 0,
\,\,\,R_{\alpha \beta \gamma \chi }\neq 0) \eqno(11)$$
\hbox to 12cm{\dotfill }

\leftline{$m\,=\,1$}
$${\bf S}_{jkl}^i\,\ =\,{\bf T}_{jkl}^i\eqno(12)$$
{$$+d\,\psi\,\cong\,\omega ^2,\qquad \omega ^2\neq 0:\quad(K_{\beta \gamma \chi }^{\alpha }
\neq 0,\,R_{\beta \gamma \chi }^{\alpha }\neq 0)\eqno(13)$$
\hbox to 12cm{\dotfill }

\leftline{$m=2$}
$${\bf S}_{kl}^{ij} \,=\,{\bf T}_{kl}^{ij}\eqno(14)$$
$$d\,\psi\,\cong\,\omega ^1,\qquad \omega ^1\neq 0:\quad(K_{\gamma \chi }^{\alpha \beta }
\neq 0,\,R_{\gamma \chi }^{\alpha \beta }\neq 0) \eqno(15)$$
\hbox to 12cm{\dotfill }

\leftline{$m=3$}
$${\bf S}_{l}^{ijk} \,=\,{\bf T}_{l}^{ijk}\eqno(16)$$
$$d\,\psi\,\cong\,\omega ^0,\qquad \omega ^0\neq 0:\quad(K_{\chi }^{\alpha
\beta \gamma }
\neq 0,\,R_{\chi }^{\alpha \beta \gamma }\neq 0)a\eqno(17)$$
\hbox to 12cm{\dotfill }

\leftline{$m=4$}
$${\bf S}^{ijkl}\,=\,{\bf T}^{ijkl}\eqno(18)$$
$$d\,\psi\,\cong\,\int \omega ^0,\qquad \omega ^0\neq 0:\quad(K^{\alpha \beta \gamma \chi }
\neq 0,\,R_{\alpha \beta \gamma \chi }\neq 0)\eqno(19)$$
\hrule
$$\psi \,=\,0$$

The line (11) in this scheme corresponds to the nonidentic initial evolutionary 
relation (with the exterior forms of degree 3). Here the inequality 
$d\,\omega ^3\neq 0$ is written in terms of the tensor expressions for the 
commutators: ($K_{\alpha \beta \gamma \chi}\neq 0$, 
$R_{\alpha \beta \gamma \chi }\neq 0$).

The dotted line corresponds to the degenerate transform and to the transition 
from the nonidentic evolutionary relation to the identic relation on the 
pseudostructure of the dimension $m=1$ (the line (12)), as well as to the nonidentic 
relation of one less degree (the line (13)). The line (12) contains 
the identic relation in the tensor expressions (see, the relation (6)), which 
corresponds to the identic relation (4) in the differential forms.

Under the degenerate transform it is once more allowed a transition from 
the nonidentic relation in the line (13) to the identic relation on the
pseudostructure of the dimension $m=2$ (the line (14)) and to the new 
nonidentic relation (the line (15)). Similar transitions can be realized 
under the degenerate transforms up to the closed inexact forms of zero 
degree. The solid line corresponds to the transition to the exact form.

A realization of the pseudostructures of dimensions $(1,\,...,\,4)$ and 
closed inexact forms of degrees $(3,\,...,\,0)$ (an origination of the physical 
structures \hbox{${\bf S}_{jkl}^i,\,...,
\,{\bf S}^{ijkl}$}) correspond to formation 
of the pseudometric manifold. A transition to the exact form corresponds to 
a transition to the metric space.

And what can one say concerning  the pseudoriemann manifold and the Riemann 
space?

As it is known, when deriving the Einstein equation [29] it was supposed that 
the following conditions to be satisfied: the Bianchi identity is fulfilled, 
the connectivity coefficients are symmetric ones (the connectivity coefficients  
are the Christoffel symbols), and there is a transformation under which 
the connectivity coefficient becomes zero. These conditions are those of 
realization of the degenerate transforms for the nonidentical evolutionary 
relations (13), (15), (17), (19). If the Bianchi identities are satisfied [21], 
then from the tensor expression $R_{\beta \gamma \chi }^{\alpha }$ 
the Riemann-Christoffel tensor $G_{jkl}^i$ can be obtained. To the tensor 
expression $R_{\gamma \chi }^{\alpha \beta }$ there corresponds the commutator 
of the first order metric form $(\Gamma _{\mu \nu }^{\rho }-\Gamma _{\nu \mu }^{\rho })$, 
from which under the conditions of symmetry of the connectivity coefficients 
$(\Gamma _{lk}^j-\Gamma _{kl}^j)=0$ the Ricci tensor can be found.
To the tensor expression $R_{\chi }^{\alpha \beta \gamma }$ there corresponds 
the connectivity $\Gamma _{\mu \nu }^{\rho }$, from which under the condition 
$\Gamma _{kl}^j=\{{j\atop kl}\}$ (the connectivity coefficients are equal to the Christoffel 
symbols) it can be obtained the tensor expression ${\bf S}_l^{ijk}$, which corresponds to 
the Einstein tensor $S_l^k=G_l^k-{1\over 2}\delta _l^kG$ (the tensors $G_l^k$ 
and $G$ are obtained from the Riemann-Christoffel tensor with taking into 
account the symmetry of the connectivity coefficients). To Einstein's equation 
there corresponds the identity (16) that connects the tensor expression 
${\bf S}_l^{ijk}$ with the tensor expression ${\bf T}_l^{ijk}$ which corresponds to 
the energy-momentum tensor. (It is well to bear in mind that the metric tensor 
has not formatted as yet, and therefore the operation of transfer of low and top subscripts 
with the help of the metric tensor proves to be inapplicable). To the tensor expression 
$R^{\alpha \beta \gamma \chi }$ there corresponds the connectivity coefficients, which 
under a presence of the degenerated transform vanish, and this corresponds to 
a formation of the closed (inexact) metric form of zero degree $g_{kl}=({\bf e}_k{\bf e}_l)$. 
However, at given stage this only corresponds to formation of the pseudoriemann 
manifold. A transition from the closed inexact form of zero degree to the exact 
form of zero degree corresponds to transition to the Riemann space.

\bigskip
\rightline{\bf Appendix 7}
\centerline{\bf Functional properties of solutions to the differential equations.}
\centerline{\bf The field equation. Transformations.}

While description of the physical processes by differential equation the 
following fact is essential. As to the exact conservation laws there have to 
correspond the closed forms, then to them may correspond only solutions, whose 
derivatives constitute the closed form, i.e. a differential. What conditions 
must the differential equations satisfy to, for to have such solutions? Let 
us trace this by the example of the first order partial differential 
equation:
$$ F(x^i,\,u,\,p_i)=0,\quad p_i\,=\,\partial u/\partial x^i \eqno(1)$$

Let us consider the functional relation
$$ du\,=\,\Theta \eqno(2)$$
where $\Theta\,=\,p_i\,dx^i$. This relation is an analog to the evolutionary
relation, and in the general case this relation (as well as the evolutionary 
relation) proves to be nonidentical. For this relation to be identical one, 
the both parts of the relation have to be  differentials, i.e. the 
closed forms. To obey this condition, the commutator 
$K_{ij}\,=\,\partial p_j/\partial x^i\,-\,\partial p_i/\partial x^j$ of the form 
$\Theta\,=\,p_idx^i$ has to be zero. However, from the equation (1) it does not evidently 
follow, that the derivatives $p_i\,=\,\partial u/\partial x^i $, which obey to the equation (and given 
boundary or initial conditions of the problem), are conjugate, that is, 
they made a commutator of the form $\Theta $ equal to zero. In the general case 
without any supplementary conditions a commutator of the form $\Theta $ is 
nonzero.The form $\Theta\,=\,p_i\,dx^i$ proves to be nonclosed and is not a 
differential unlike the left-hand side of the relation (2). The functional 
relation (without supplementary conditions) proves to be nonidentical. And 
as the derivatives of the initial equation do not format a differential, 
then the corresponding solution to the differential equation $u$ will not be 
a function of $x^i$. It will depend on the commutator of the form $\Theta $, 
that is, it will be a functional.

To obtain the solution, which is the function, it is necessary to add a closure 
condition for the form $\Theta\,=\,p_idx^i$ and for the form dual to that 
(in the present case the functional $F$ plays a role of the form dual to 
$\Theta $) [3]:
$$\cases {dF(x^i,\,u,\,p_i)\,=\,0\cr
d(p_i\,dx^i)\,=\,0\cr}\eqno(3)$$
If expand the differentials, then we get a set of the homogeneous equations 
with respect to $dx^i$ and $dp_i$ (in space of the dimension $2n$ -- initial 
and tangential):
$$\cases { \left ({{\partial F}\over {\partial x^i}}\,+\,
{{\partial F}\over {\partial u}}\,p_i\right )\,dx^i\,+\,
{{\partial F}\over {\partial p_i}}\,dp_i \,=\,0\cr
dp_i\,dx^i\,-\,dx^i\,dp_i\,=\,0\cr} \eqno(4)$$ 
The solvability conditions for this set (a vanishing of the determinant 
composed of the coefficients at $dx^i$, $dp_i$) have the form:
$$ 
{{dx^i}\over {\partial F/\partial p_i}}\,=\,{{-dp_i}\over {\partial F/\partial x^i+p_i\partial F/\partial u}} \eqno (5)
$$
These conditions determine an integrating direction, namely, a pseudostructure, 
on which the form $\Theta \,=\,p_i\,dx^i$ turns out to be closed one, i.e. it becomes 
a differential, and the relation (2) proves to be identity. If the conditions (5), 
which may be called the integrability conditions, are fulfilled, the derivatives constitute 
a differential ($\delta u\,=\,p_idx^i\,=\,du$) and the solution becomes the 
function. Just such solutions (functions on the pseudostructures) are 
so-called generalizated solutions [30]. $\{$As the functions, which are the 
generalizated solutions (distributions), 
are defined only on the pseudostructures, then they have breaks of derivatives 
in directions being transverse to the pseudostructures. The order of derivatives, 
which have breaks, is equal to a degree of the exterior form. If the form 
of zero degree enters into the functional relation, then the function itself 
will have the breaks$\}$.

If to find the characteristics of the equation, then it appears that the conditions (5) 
are the equations for characteristics [31]. The characteristics are the pseudostructures, 
on which the solutions prove to be functions (generalized solutions). $\{$The 
coordinates of the equations for characteristics are not identical to the independent 
coordinates in the initial equation (1). A transition from coordinates of the initial 
space to the characteristic manifold appears to be the {\bf degenerate} transform, namely, 
the determinant of the set of equations (4) becomes zero. The derivatives of the 
equation (1) transfer from the tangent space to cotangent one$\}$.

A partial differential equation of the first order has been analyzed, 
and the functional relation with the form of the first degree has been considered. 
(At this point it worth noting that for this equation one has to write down 
and analyze the additional relation with the zero-order form as well). Similar 
functional properties have the solutions to all differential equations. And, if 
the order of the differential equation is $k$, then to this equation there 
corresponds 
$(k+1)$ functional relations, every of which contains the exterior forms of 
degrees: $k,\,k-1,...,\,0$. In a similar manner one can investigate the solutions 
to a set of the differential equations and the ordinary differential equations 
(for which the nonconjugativity of the unknown functions and initial 
conditions are examined). $\{$As it is known, the analyses of the unstable solutions 
and the integrability conditions provides the basis of the qualitative theory of the 
differential equations. From the functional relation it follows that to the 
instability it leads a dependence of the solution on the commutator, and as the 
integrability condition it serves the closure conditions of the form composed 
of derivatives. It is evident that an analyses of the nonidentical functional 
relation lies at the basis of the qualitative theory of the differential 
equations$\}$.

The functional properties of the differential equation play an essential 
role under description of the physical processes. It is clear, that to the 
conservation laws, and hence, to the physical structures, there can correspond 
only generalized solutions, whose derivatives format the closed form. The 
solutions-functionals have a physical sense as well. The solutions to the 
equations of the balance conservation laws, which are functionals, describe 
nonequilibrum states of the material system. And a transition (under the 
degenerate transformation) from the solution-functional to the generalized 
solution (a transition from the nonintegrable form to the closed one) 
corresponds to origin of the physical structure. 
 
Because of that to the physical structures, which format the physical fields, 
there correspond the closed forms, then only the equations with the additional 
conditions (integrability conditions) can be the equations of field theory. 

Let us return to the equation (1). Suppose, that it does not explicitly 
depend on $u$ and it is solved with respect to some variable, for example 
$t$, that is, it has the form
$${{\partial u}\over {\partial t}}\,+\,E(t,\,x^j,\,p_j)\,=\,0, \quad p_j\,=\,{{\partial u}\over {\partial x^j}}\eqno(6)
$$
Then the integrability conditions (5) take the form (in this case 
$\partial F/\partial p_1=1$)
$${{dx^j}\over {dt}}\,=\,{{\partial E}\over {\partial p_j}}, \quad 
{{dp_j}\over {dt}}\,=\,-{{\partial E}\over {\partial x^j}}\eqno(7)$$

The conditions (7) are known as the canonical relations. (It can be seen 
that the canonical relations are the equations for characteristics of the 
equation (6)). The equation (6) provided with the supplementary conditions, 
namely, the canonical relations (7), is called the Hamilton-Jacobi equation [31]. 
The derivatives of this equation format the differential:
$\delta u\,=\,(\partial u/\partial t)\,dt+p_j\,dx^j\,=\,-E\,dt+p_j\,dx^j\,=\,du$.
To this type there belongs the equation of field theory
$${{\partial s}\over {\partial t}}+H \left(t,\,q_j,\,{{\partial s}\over {\partial q_j}}
\right )\,=\,0,\quad 
{{\partial s}\over {\partial q_j}}\,=\,p_j \eqno(8)$$
where $s$ is the field function for the action functional $S\,=\,\int\,L\,dt$. 
Here $L$ is the Lagrange function, $H$ is the Hamilton function: 
$H(t,\,q_j,\,p_j)\,=\,p_j\,\dot q_j-L$, $p_j\,=\,\partial L/\partial \dot q_j$. 
To the equation (8) there correspond the closed form $ds\,=\,H\,dt\,+\,p_j\,dq_j$ (the 
Poincare invariant). $\{$ In the quantum theory an analog to the equation (8) 
is the Schr\H{o}dinger equation [32]$\}$.

$\{$Here the degenerate transformation is a transition from the Lagrange 
function to the Hamilton function. An equation for the Lagrange function, that 
is the Euler variational equation, 
has been obtained from the condition $\delta S\,=\,0$, where $S$ is the action 
functional. In the real case, when forces are nonpotential or connections are 
nonholonomic, the quantity $\delta S$ is not a closed form, that is, $d\,\delta S\,\neq \,0$.
But the Hamilton function is obtained from the condition $d\,\delta S\,=\,0$ 
which is the closure condition for the form $\delta S$. A transition from the 
Lagrange function $L$ to the Hamilton function $H$ (a transition from 
variables $q_j,\,\dot q_j$ to variables $q_j,\,p_j=\partial L/\partial \dot q_j$) is 
a transition from the tangent space, where the form is nonclosed, to 
cotangent space with closed form. One can see, that this transition is 
degenerate one$\}$. $\{$In the invariant field theories there used only 
nondegenerate transformations, which preserve the differential. 
By the example of the canonical relations it is possible to show that 
nondegenerate and degenerate transformations are connected. The canonical relations 
in the invariant field theory correspond to nondegenerate tangent transformations. 
At the same time, the canonical relations for the Hamilton-Jacobi equation 
without supplementary conditions are the equations for characteristics, which 
the degenerate transformations correspond to. The degenerate transformation is 
a transition from the tangent space ($q_j,\,\dot q_j)$) to the cotangent (characteristic) 
manifold ($q_j,\,p_j$). It is a transition from manifold that corresponds to the 
material system, to the physical fields, and it is an origin of the physical 
structure. On the other hand, the nondegenerate transformation is a transition 
from one characteristic manifold ($q_j,\,p_j$) to the other characteristic 
manifold ($Q_j,\,P_j$), that is, a transition from one physical structure 
to another physical structure. It is easily shown that it is a specific 
feature for the relations, which perform such transformations as tangent, 
gradient, contact, gauge, conformal mapping, and others$\}$.

\bigskip
\centerline{References}

1. Wheeler J.~A., Neutrino, Gravitation and Geometry. Bologna, 1960.

2. Schutz B.~F., Geometrical Methods of Mathematical Physics. Cambrige 
University Press, Cambrige, 1982.

3. Cartan E. Les Systemes Differentials Exterieus ef Leurs Application 
Geometriques. Hermann, Paris, 1945.  

4. Encyclopedic dictionary of the physical sciences. -Moscow, Sov.~Encyc., 
1984 (in Russian).

5. Petrova L.~I., Exterior differential forms in the field theory. 
//Abstracts of International Conference dedicated to 
the 90$^{th}$ Anniversary of L.~S.~Pontryagin, 
Algebra, Geometry, and Topology. Moscow, 1998, 123-125.

6. Petrova L.~I., A mechanism of structure formation  and the field selforganization. 
//Proc. Internal Symposium "Generation of Large-Scale Structures in Continuous 
Media" (The Nonlinear Dynamics of Structures). Perm - Moscow, 1990, 201-202.

7. Petrova L.~I., On the problem of development of the flow instability. //Basic 
Problems of Physics of Shock Waves. Chernogolovka, 1987, V.~{\bf 1}, P.~1, 304-306. 

8. Petrova L.~I., The evolutionary properties of the conservation laws. //Abstracts of 
reports at the 27$^{th}$ Science Conference on the Mathematical Physics. Edition by 
the Unuversity of the People Friendship, Moscow, 1995.

9. Petrova L.~I., Exterior differential forms in field theory. //Abstracts of 
reports at 10$^{th}$ Russian Conference on Gravitation "Theoretic and Experimental 
Problems on General Theory of Relativity and Gravitation", Moscow, 1999, 65.

10. Petrova L.~I., A role of the balance and exact conservation laws 
in origin of the physical structures. //Proc. of International 
School-Seminar "Nonlinear Problems of Theory of the Hydrodynamic Instability 
and Turbulence". Moscow, 1998, 133-136. 

11. Bott R., Tu L.~W., Differential Forms in Algebraic Topology. Springer, NY, 1982.
 
12. Encyclopedia of Mathematics. -Moscow, Sov.~Encyc., 1979 (in Russian).

13. Novikov S.~P., Fomenko A.~P., Elements of the differential geometry and 
topology. -Moscow, Nauka, 1987 (in Russian). 
 
14. Konopleva N.~P. and Popov V.~N., The gauge fields. Moscow, Atomizdat, 1980 
(in Russian).

15. Fock V.~A., Theory of space, time, and gravitation. -Moscow, 
Tech.~Theor.~Lit., 1955 
(in Russian).

16. Clark J.~F., Machesney ~M., The Dynamics of Real Gases. Butterworths, 
London, 1964.

17. Tolman R.~C., Relativity, Thermodynamics, and Cosmology. Clarendon Press, 
Oxford,  UK, 1969.

18. Dafermos C.~M. In "Nonlinear waves". Cornell University Press, 
Ithaca-London, 1974.

19. Glansdorff P., Prigogine I. Thermodynamic Theory of Structure, Stability 
and Fluctuations. Wiley, N.Y., 1971.  

20. Rumer Yu.~B. Investigations on 5-Optics. GITTL, Moscow, 1956 (in Russian).

21. Tonnelat M.-A., Les principles et de la theorie electromagnetique et la relattivete. 
Masson, Paris, 1959.

22. Pauli W. Theory of Relativity. Pergamon Press, 1958.

23. Haywood R.~W., Equilibrium Thermodynamics. Wiley Inc. 1980.

24. Petrova L.~I., A mechanism of development of gas dynamic instability 
and of origin of the gas dynamic structures. //Proc of the 1$^{st}$ Russian 
Conference on Heat Transfer, Moscow, 1994, V.~{\bf 1}, 201-206.

25. Liepman H.~W., Roshko ~A., Elements of Gas Dynamics. Jonn Wiley, 
New York, 1957.

26. Dirac P.~A.~M., Proc.~Roy.~Soc., {\bf A133}, 60 (1931).

27. Dirac P.~A.~M., Phys.~Rev., {\bf 74}, 817 (1948).

28. Weinberg S., Gravitation and Cosmology. Principles and applications of 
the general theory of relativity. Wiley \& Sons, Inc., N-Y, 1972.

29. Einstein A. The Meaning of Relativity. Princeton, 1953.
 
30. Vladimirov V.~S., Equations of the mathematical physics. -Moscow, 
Nauka, 1988 (in Russian).

31. Smirnov V.~I., A cource of higher mathematics. -Moscow, 
Tech.~Theor.~Lit. 1957, V.~4.

32. Dirac P.~A.~M., The Principles of Quantum Mechanics. Clarendon Press, 
Oxford, UK, 1958.

\end{document}